\pdfoutput=1

\documentclass[twocolumn]{emulateapj}

\usepackage{graphicx}
\usepackage{amsmath, amsthm, amssymb}
\usepackage{natbib}
\usepackage{epsfig}
\usepackage{subfigure}
\usepackage[colorlinks = true, linkcolor=blue, citecolor=blue, urlcolor=blue]{hyperref}
\usepackage{color}
\usepackage[CaptionAfterwards]{fltpage}

\usepackage{float}

\newcommand{\Enzo}{\texttt{Enzo}}
\newcommand{\yt}{\texttt{yt}}
\newcommand{\Mpch}{$h^{-1}$ Mpc}
\newcommand{\kpch}{$h^{-1}$ kpc}
\newcommand{\cloudy}{Cloudy}
\newcommand{\hif}{\textit{HIFLUGCS}}

\defcitealias{2008ApJ...675.1125B}{B08}
\defcitealias{2010A&A...513A..37H}{H10}

\shorttitle{On The Road To More Realistic Clusters}
\shortauthors{Skory et al.}

\begin{document}


\title{On The Road To More Realistic Galaxy Cluster Simulations: 
    The Effects of Radiative Cooling and Thermal Feedback Prescriptions on
    the Observational Properties of Simulated Galaxy Clusters}


\author{Stephen Skory$^{\dagger}$, Eric Hallman\altaffilmark{a}, Jack O. Burns,
	Samuel W. Skillman\altaffilmark{b}}
\affil{Center for Astrophysics \& Space Astronomy, Department of Astrophysical \& Planetary Sciences,
	389 UCB, University of Colorado, Boulder, CO 80309, USA}
\email{$^{\dagger}$stephen.skory@colorado.edu}
\and
\author{Brian W. O'Shea\altaffilmark{c}, Britton D. Smith}
\affil{Department of Physics \& Astronomy, Michigan State University,
	East Lansing, MI 48824, USA}

\altaffiltext{a}{Also at Tech-X Corporation, Boulder, CO, 80303, USA}
\altaffiltext{b}{DOE Computational Science Graduate Fellow}
\altaffiltext{c}{Also at Lyman Briggs College and Institute for Cyber-Enabled Research,
Michigan State University, East Lansing, MI 48824, USA}


\begin{abstract}
Flux limited X-ray surveys of galaxy clusters show that
clusters come in two roughly equally proportioned varieties:
``cool core'' clusters (CCs) and non-``cool core'' clusters (NCCs).
In previous work, we have demonstrated using cosmological $N$-body + Eulerian hydrodynamic simulations 
that NCCs are often consistent with early major mergers events that destroy embryonic CCs.
In this paper we extend those results and conduct a series of simulations
using different methods of gas cooling, and of energy and metal feedback from supernovae,
where we attempt to produce a population of clusters with realistic
central cooling times, entropies, and temperatures.
We find that the use of metallicity-dependent gas cooling is essential to prevent early overcooling,
and that adjusting the amount of energy and metal feedback can have a significant impact on
observable X-ray quantities of the gas.
We are able to produce clusters with more realistic
central observable quantities than have previously been attained.
However, there are still significant discrepancies between the simulated
clusters and observations, which indicates that a different approach to simulating galaxies in clusters
is needed.
We conclude by looking towards a promising subgrid method of modeling
galaxy feedback in clusters which
may help to ameliorate the discrepancies between simulations and observations.
\end{abstract}



\section{Introduction}

A wide variety of X-ray surveys
\citep[e.g.,][]{2006MNRAS.372.1496S, 2007A&A...466..805C, 2009MNRAS.395.1287J}
of galaxy clusters give the interesting result that the proportion of ``cool core'' (CC) clusters and non-``cool core'' (NCC) clusters 
is roughly equal.
While authors use different specific definitions for CC clusters,
they are fundamentally differentiated from NCC clusters by bright, cuspy central X-ray profiles
indicative of high central densities and correspondingly short cooling times ($<H_0^{-1}$),
and low central temperatures
\citep[$\approx30\%$--$40\%$ of the virial temperature; ][]{1997ApJ...481..660I, 2002ApJ...573L..13L, 2003ApJ...590..207P}.
Early observations of CC clusters \citep{1973ApJ...184L.105L, 1977ApJ...215..723C,
1977MNRAS.180..479F, 1978ApJ...224..308M} motivated the creation of the
``cooling flow'' model (see \citet{1994ARA&A..32..277F} for a review)
to explain the existence of CCs within galaxy clusters, wherein central gas quasi-hydrostatically cooled
to low temperatures, and was replenished with hot gas
flowing in from the intracluster medium (ICM).
However, the cooling flow model suggests a higher star formation rate than is observed
\citep{1989AJ.....98.2018M, 1998MNRAS.298..977C, 2001MNRAS.328..762E},
predicts a high mass deposition rate
\citep{2001PASJ...53..401M},
and cooler central temperatures
\citep{2001A&A...365L.104P, 2008MNRAS.385.1186S},
and therefore newer ideas have been created to explain the bimodality.

Theories to explain the discrepancies often include additional sources and methods of transport of heat in a cluster.
These ideas include thermal heat conduction \citep{2003ApJ...582..162Z, 2011ApJ...740...28V}
combined with sound waves \citep{2004ApJ...611..158R}
or with turbulence \citep{2005ApJ...622..205D},
conduction combined with cosmic rays \citep{1991ApJ...377..392L, 2008MNRAS.384..251G},
and active galactic nuclei (AGN) \citep{1995MNRAS.276..663B, 1995ApJ...442...91R, 2007ARA&A..45..117M}.
Numerical simulations have demonstrated that AGN may be important sources of feedback energy
and significantly impact the energy balance in clusters.
Simulations show that AGN bring simulations closer to observations
by preventing overcooling, which lowers cluster central densities
\citep{2007MNRAS.380..877S, 2011MNRAS.414..195T},
improves the distribution of metals \citep{2010MNRAS.401.1670F, 2010MNRAS.406..822M},
and regulates star formation rates and triggers quenching \citep{2005Natur.433..604D, 2006MNRAS.366..397S}.

Other phenomena that may explain the observed cluster population properties include those
methods that disturb the gas at the centers of clusters, or which disrupt the cool core.
This class of phenomena is most naturally explained by merger events between halos.
It is, of course, impossible to observationally follow cluster mergers from start to finish and measure its effect on cluster cores,
but there is some evidence that NCC clusters are associated with past merger events.
By measuring the offset between the brightest cluster galaxy and the X-ray centroid,
\citet{2009MNRAS.398.1698S} find that clusters with larger offsets,
which imply a state of dynamical disturbance, have weaker CCs.
In their sample, \citet{2011A&A...532A.123R} find no clusters with giant radio halos 
(which are associated with past merger events) that can be classified as CC.
Using detailed density profiles of observed CC and NCC clusters,
\citet{2012A&A...541A..57E} find that as compared to CC,
the outskirts of NCC clusters have flatter profiles, which they argue is evidence of past major merger
events which redistribute gas between all regions of the cluster more effectively than 
centrally-dominated sources of entropy, such as AGN.
\citet{2009ApJ...697.1597H} find that observed NCC clusters are warmer in the periphery than CC clusters,
and that an analogous population of NCC clusters sampled from numerical simulations have a richer history of
early merger events compared to their simulated CC cluster counterparts.

Previously, our group has used cosmological simulations to explore the idea that mergers influence the creation of NCC clusters.
In \citet{2008ApJ...675.1125B} (hereafter \citetalias{2008ApJ...675.1125B})
we performed simulations that included gas cooling plus star formation with supernovae feedback, and we found that
galaxy clusters are born cool, but may warm up when impacted by a
major merger at some point in their early histories.
Interestingly, \citet{2010A&A...510A..83R} find most of their sampled observed NCC
clusters have regions of high metallicity and low entropy gas,
which they interpret as indicators of a past cool phase for NCC clusters.
Using simulations, \citet{2010ApJ...717..908Z} show that when the core of a cluster interacts
with a merging subcluster, the core gas can enter a ``sloshing'' phase that
can delay the development of the cool core for one or more Gyr.
In a related paper, \citet{2011ApJ...728...54Z} finds that cluster mergers over a
wide range of mass ratios and impact parameters act to mix up the phases of gas
and raise the overall entropy floor, making CC formation (or preservation) more difficult.

The assumption that the galaxy cluster population is approximately evenly split between cool core
and non-cool core clusters is likely too simplistic.
It is probable that there is an observational bias in the CC/NCC cluster ratio in surveys because
flux-limited X-ray observations will naturally over-sample the centrally brighter CC groups.
Observations by \citet{2011A&A...526A..79E} suggest that there is a
bias towards CC clusters of as much as 30\%.
\citet{2010A&A...513A..37H} (hereafter \citetalias{2010A&A...513A..37H}) used
\hif\ (HIghest X-ray FLUx Galaxy Clusters Sample)
data \citep{2002ApJ...567..716R} to do a statistical analysis on a large number
of cluster measureables in an attempt to establish cutlines between
the CC and NCC subgroups.
For a few of the observables, most notably central cooling time and entropy (see their Fig. 4),
they find that the best statistical fit in fact divides the population into three groups:
NCCs, Weak CCs (WCCs), and Strong CCs (SCCs).
We discuss the work presented in \citetalias{2010A&A...513A..37H} in more detail in \S\ref{sec:ObsData}.

In \citetalias{2008ApJ...675.1125B} we defined a CC cluster as one with a $\geq20\%$
drop in central gas temperature compared to the surrounding gas.
Using this definition, we produced both CC and NCC clusters in a single
cosmological simulation for the first time.
The work presented in this paper builds on the results of
\citetalias{2008ApJ...675.1125B} in two ways.
First, {\it we focus on comparing our results with observations to understand the effects of the different physical models},
and use these comparisons to modify our simulations in an effort to make more realistic cluster populations.
Second, {\it we include two physically-motivated numerical models}, one for gas cooling and
one for metal and energy feedback, that we use in an attempt to address shortcomings in our previous work.
We compare and contrast our simulations in order to understand how the physical models affect the clusters,
in a way inspired by the pioneering work of \citet{2000ApJ...536..623L}
that examined in detail the effect of gas cooling on simulated galaxy clusters.
Many groups have investigated the simulation of clusters with gas cooling and supernovae feedback
\citep[e.g.][]{2003astro.ph.10203V, 2004Ap&SS.294...51B, 2005ApJ...625..588K, 2007MNRAS.377..317K,
2007ApJ...668....1N, 2007MNRAS.382.1050T, 2011MNRAS.416..801F},
but in this paper, we use similar physics - however we use a large set of clusters ($\sim65-70$ per simulation)
and focus on only the properties of core of the clusters.
As we will show, when additional cluster observables are considered,
the \citetalias{2008ApJ...675.1125B} simulations
fail to reproduce other observed cluster characteristics such as the central entropy.
Therefore, our motivations for including the new models of cooling and feedback
are to address clear deficits in the methods used in \citetalias{2008ApJ...675.1125B},
improve the concordance with observations for our simulated clusters,
and discover the shortcomings of current methods and how they might motivate future numerical models.

We note that the simulations discussed in this paper do not include a prescription for AGN formation and feedback.
We have already listed some of the benefits of AGN in cluster simulations,
and it is well known that simulations that include only cooling + star formation
do not do a good job of reproducing all of the observed properties of clusters
\citep{2006MNRAS.366..397S, 2007MNRAS.380..877S, 2008ApJ...687L..53P}.
This paper is, therefore, an exploration of the limitations of models that use only stellar feedback,
and quantifies how well (or poorly) this type of simulation performs when compared against
standard cluster observable quantities.

In Section \ref{sec:NumSim} we describe our simulations and numerical methods,
including the details of the gas cooling and feedback mechanisms.
We base our comparisons to observations using the results presented in
\citetalias{2010A&A...513A..37H} of \hif\ data,
which we describe in detail in \S\ref{sec:ObsData}.
Our first four simulations, presented in \S\ref{sec:FirstFourSims},
investigate the effects of a variety of methods for gas cooling and stellar feedback
between simulations, and we also compare to observations.
In response to the results of \S\ref{sec:FirstFourSims}, we perform three additional simulations
and compare them to observations in \S\ref{sec:AdjustFeedback}.
In \S\ref{sec:Discussion} we summarize the results and shortcomings of our simulations, and
in \S\ref{sec:Con} we present our conclusions and a discussion regarding the steps
that must be taken in order to produce more realistic clusters.

\section{Numerical Methods}\label{sec:NumSim}

Our simulations are performed using \Enzo\footnote{\url{http://enzo-project.org/};
our simulations were run using the \Enzo\ development branch with Mercurial revision identifier \texttt{d12ea971621c}.}
\citep{Enzo2005, Enzo2007}, an open-source, community-developed code that solves
hydrodynamics on an Eulerian mesh.
In these simulations, we use the ZEUS scheme for solving hydrodynamics \citep{1992ApJS...80..753S},
with an \it N\rm-body scheme for evolving the dark matter and star particles.
\Enzo\ includes adaptive mesh refinement (AMR) capability, which we discuss in more detail below.
\Enzo\ has been compared to other cosmological codes with favorable results for ICM physics
\citep{1999ApJ...525..554F, 2005ApJS..160....1O, 2007MNRAS.380..963A, 2008MNRAS.390.1267T, 2011MNRAS.418..960V}.
We use a $\Lambda$CDM cosmology, and our initial conditions are
generated at $z = 99$ using the CDM transfer function from
\citet{1999ApJ...511....5E} using the parameters $\Omega_m=0.268$,
$\Omega_{\Lambda}=0.732$, $\Omega_b = 0.0441$, h$= 0.704$,
$\sigma_8 = 0.82$, and $n_s = 0.97$ \citep[WMAP3;][]{2007ApJS..170..377S}.

Our simulations employ a $128^3$ (\Mpch)$^3$ volume with $256^3$ top-level
grid zones and dark matter particles.
Dark matter particles are given a mass of $7.8\times10^9$ h$^{-1}$M$_\odot$,
and the mean baryon mass of a root-grid cell is $3.1\times10^9$ h$^{-1}$M$_\odot$.
AMR is enabled throughout the entire volume on cells containing baryonic and/or
dark matter density 8.0 times the mean at that level, for up to 5 additional levels.
This results in a peak co-moving spatial resolution of 15.6 \kpch, which is sufficient to resolve
features having a scale of $\sim50-100$ \kpch\ such as the cool cores of clusters,
but is insufficient to resolve any more detailed structure.

\subsection{Cooling Physics}\label{sec:cooling}

Radiative cooling of gas is applied on every cell at every time step.
Using one of the two cooling methods described below,
an amount of energy to be radiated away is calculated,
and that energy is subtracted from the cell.
For all simulations using the metal-dependent \cloudy\ cooling method (described below),
a uniform, metagalactic ionizing UV background \citep[$q_{\alpha}=1.5$,][]{1996ApJ...461...20H} from quasars is applied
after z=7.

Two of our simulations use metallicity-independent Raymond-Smith cooling \citep[hereafter ``RS cooling''; ][]{1976ApJ...204..290R},
the identical method to the one used in the \citetalias{2008ApJ...675.1125B} simulations.
RS cooling rates are computed from an optically thin fully-ionized plasma emission model
\citep{1995ApJS...97..551B, 1987ApJ...320...32S} that assumes a {\it constant metallicity of 0.5 relative to solar ($Z_{\odot}$)}.
Analytical approximations of cooling rates are stored in a lookup table as a function of temperature,
and the cooling rate of a cell is found by multiplying by the density squared of the cell.
We have run a simulation with RS cooling using rate tables computed assuming a constant $Z/Z_{\odot}=0.25$ metallicity
and did not find any significant changes from $Z/Z_{\odot}=0.5$.
Using a cooling function dependent only on temperature is simplistic -- in particular,
the assumption of constant metallicity is questionable.
As we will show in Section \ref{sec:ABCloudy}, the assumption of a constant, high metallicity leads to overcooling of gas at
high redshift and substantially affects cluster observables.

To address this issue, we replace RS cooling with a cooling model partially based on the
photoionization code \cloudy\ \citep{1998PASP..110..761F}.
Throughout this paper, we use the term ``\cloudy\ cooling'' to describe this
model that includes {\it metallicity-dependent} cooling rates \citep{2008MNRAS.385.1443S, 2011ApJ...731....6S}.
The simulations that utilize \cloudy\ cooling
actually combine two methods to calculate cooling rates,
and the total rate is the sum of the two.
For atomic H and He, the individual abundances are tracked and the non-equilibrium cooling
rates are computed directly via a network of coupled equations \citep{1997NewA....2..181A, 1997NewA....2..209A}.
The metal cooling rates from all atomic species between Li and Zn (assuming solar abundances for relative species fractions),
and many molecular species as well,
are found by referencing pre-computed tables that depend on
temperature, density, electron fraction, and (crucially) metallicity up to z=7,
and also the UV background after z=7.
The tables are built by using the \cloudy\ code to compute the equilibrium rates of
cooling over a grid of values in the phase space of the aforementioned dependencies,
and the rates are interpolated from the grid of values by \Enzo\ as the simulation runs.
The range of the dependencies used to build the tables covers all the relevant physical values for our simulations.

\subsection{Star Formation and Feedback}\label{sec:stars_feed}

Just as radiative cooling is an important part of the energetics of a galaxy cluster,
so is star formation and the associated feedback.
As in \citetalias{2008ApJ...675.1125B}, we include a prescription for star formation in our simulation
using the formulation described in  \citet{1992ApJ...399L.113C},
which we briefly describe here.
In all grid cells that are locally at the most highly-refined level,
a collisionless ``star particle'' is created in a cell if the following conditions are met in that cell:
1) the gas density is at least 100 times higher than the mean gas density in the volume;
2) the flow of gas is locally converging;
3) the gas cooling time is less than the total (gas + dark matter) dynamical time;
and, 4) and the mass of gas in the cell exceeds the Jean's mass.
The mass of the star $M_{\mathrm{star}} = M_{\mathrm{cell}} \Delta t / t_{\mathrm{dyn}}$
depends on the mass of gas in the cell $M_{\mathrm{cell}}$,
the length of the time step $\Delta t$,
and the dynamical time of the gas in the cell $t_{\mathrm{dyn}}$, which is calculated for every cell and has a minimum of 1 Myr.
A star particle is formed only if the minimum mass constraint is satisfied (due to computational limitations),
which we set to $M_{\mathrm{star}} \geq 10^9$ M$_{\odot}$.
The star particle is given a metallicity $Z_{\mathrm{star}}$
where the value is same as the metallicity of the gas from which the particle is formed.

After the star is formed, energy is deposited back into the gas
to simulate feedback from Type II supernovae.
The amount of energy feedback

\begin{equation}\label{eqn:E}
e = \epsilon_E M_{\mathrm{star}} c^2
\end{equation}
is controlled by the dimensionless input parameter $\epsilon_E \ll 1$.
In all our simulations star particles also return metals to the gas in order to model
enrichment from supernovae.
Recall that RS cooling assumes a constant metallicity, and although
star particles increase the metallicity of the surrounding gas in simulations employing RS cooling,
the calculated cooling rates are completely independent of this effect.
However, we will show that gas metallicity is very important in calculations that use the metallicity-dependent
\cloudy\ cooling model.
The total mass of metals $M_Z$ returned from the star particle to the gas phase is described by the equation

\begin{equation}\label{eqn:Z}
M_{Z} = M_{\mathrm{star}} \left[(1 - Z_{\mathrm{star}}) \epsilon_Z + 0.25Z_{\mathrm{star}}\right],
\end{equation}
where $\epsilon_Z < 1$ is a dimensionless input to our simulations.
Both the energy and metal feedback is applied over a $12t_{\mathrm{dyn}}$
time period after the particle is formed, where the rates rise linearly for $t < t_{\mathrm{dyn}}$
and then decrease exponentially after $t_{\mathrm{dyn}}$.
The total feedback quantities are independent of simulation time step, but the increased
heat in the gas may shorten the time steps due to the Courant condition.
The value of $t_{\mathrm{dyn}}$ varies for each star particle (it is assigned at creation),
and in the cluster simulations discussed in this paper
it ranges from roughly 10 to 30 Myr.
See Table \ref{table:SimParams} for the values of $\epsilon_E$ and $\epsilon_Z$
used in each of our simulations.

\begin{deluxetable*}{rccccc}
\tabletypesize{\scriptsize}
\tablecaption{Simulation Parameters\label{table:SimParams}}
\tablewidth{0pt}
\tablehead{
\colhead{Simulation Label} & \colhead{Cooling} & \colhead{\# of Feedback Cells} &
\colhead{$\epsilon_E$\tablenotemark{a}} & \colhead{$\epsilon_Z$\tablenotemark{b}} &
\colhead{\# of Clusters}
}
\startdata
RS-Single & RS & 1 & $1.0\times10^{-5}$ & 0.1 & 78 \\
RS-Dist & RS & 27 &$1.0\times10^{-5}$ & 0.1 & 79 \\
Cloudy-Single & \cloudy & 1 & $1.0\times10^{-5}$ & 0.1 & 67 \\
Cloudy-HZ-LE & \cloudy & 27 &$1.0\times10^{-5}$ & 0.1 & 78 \\
\noalign{\smallskip}
\hline
\noalign{\smallskip}
Cloudy-LZ-LE & \cloudy & 27 & $1.0\times10^{-5}$ & 0.02 & 79 \\
Cloudy-LZ-ME & \cloudy & 27 & $8.0\times10^{-5}$ & 0.02 & 76 \\
Cloudy-LZ-HE & \cloudy & 27 & $2.0\times10^{-4}$ & 0.02 & 71 \\
Adiabatic & N/A & N/A & N/A & N/A & 80
\enddata
\tablenotetext{a}{Dimensionless energy feedback parameter;
	$e = \epsilon_E M_{\mathrm{star}} c^2$.}
\tablenotetext{b}{Dimensionless metallicity feedback parameter;
	$M_{Z} = M_{\mathrm{star}} ((1 - Z_{\mathrm{star}}) \epsilon_Z + 0.25Z_{\mathrm{star}})$.}
\end{deluxetable*}

In most grid-based simulations that include star formation with feedback (including those in \citetalias{2008ApJ...675.1125B}),
energy and metals are deposited entirely within the cell that contains a particular star.
As discussed in \citet{2011ApJ...731....6S}, dumping all of the thermal or kinetic
feedback into a single high density cell may result in calculated cooling times significantly shorter than a hydrodynamical time step,
which results in overcooling the gas in the cell.
In particular, overcooled gas prevents the stellar feedback from being spread out into the intracluster medium.
In order to address this,
\citeauthor{2011ApJ...731....6S} implemented in \Enzo\ the ``distributed feedback'' method,
wherein energy and metal feedback are deposited into more than one cell surrounding the star particle.
The same total quantity of feedback is injected into the gas regardless of the number of cells,
and each cell receives an equal share.
\citeauthor{2011ApJ...731....6S} show (e.g., their Fig. 2) that simulations that use \cloudy\ cooling + distributed feedback produce
star formation rates that more closely resemble the mean cosmic star formation history than those that use
single-cell feedback.
For all of our simulations that use it (see Table \ref{table:SimParams}),
distributed feedback is applied over a $3\times3\times3$ cube of cells at the highest resolution
centered on the cell containing the star.

\subsection{Analysis Methods}\label{sec:Analysis}

All simulations are evolved to and analyzed at z=0.
Our analyses are performed using the \yt\footnote{\url{http://yt-project.org/}}
toolkit \citep{2011ApJS..192....9T}.
Cluster dark matter halos are located using a parallel version of the HOP halo finder \citep{2010ApJS..191...43S},
and only virialized clusters with total (gas + dark matter) mass M$_{200}$ greater than $10^{14}$ M$_{\odot}$
are kept for our samples.
We define M$_{200}$ as the mass enclosed inside a sphere centered on the cluster with
average density 200 times the mean background density.

Central quantities and radial profiles of observed clusters are
typically found first by centering on the peak of X-ray emission.
X-ray emission is roughly proportional to $\rho^2 T^{1/2}$,
therefore we center our profiles on \it a\rm\ gas density peak close to the center of the cluster,
but not necessarily \it the absolute\rm\ gas density peak, for the following reason.
In many of our simulated clusters, infalling clumps of gas are dense enough such that their
cold centers are not shock heated immediately as they pass through the outer virialized gas of the cluster.
This means that the absolute point of maximum gas density can be offset from the actual central X-ray peak of the cluster.
The proper center of a cluster can be located by eye.
However, this is tedious and slow, so we have developed an automated method that in our tests is nearly always
(better than 99\%) identical to manual identification.

First, we find gas density ``clumps'' inside the cluster defined by regions of gas of at least
0.025 times the maximum gas density in the cluster.
Within a factor of a few, the results are not very sensitive to the choice of density threshold,
but values much higher or lower will result in incorrect identification of cluster centers.
The clumps defined by the threshold are not necessarily gravitationally bound;
they are simply defined by the volume enclosed by the density contour.
Next, we identify the location and value of maximum density inside each clump.
For smooth clusters, the point of absolute maximum density is coincident with the maximum inside the single clump.
For clusters with significant substructure, however, there are multiple discrete clumps,
and each has its own maximum density point.
To choose the proper X-ray peak, we pick the maximum density point of the clump that minimizes the equation

\begin{equation}
C = \left(\frac{R}{R-d}\right) \frac{\left|\mathbf{v}_{\mathrm{cluster}} - \mathbf{v}_{\mathrm{clump}} \right|^2}{\left| \mathbf{v}_{\mathrm{cluster}}\right| } \left(\frac{M_{\rm cluster}}{M_{\rm clump}}\right),
\end{equation}
where $R$ is the radius of the cluster, $d$ is the distance from the center of dark matter mass of the cluster
to the maximum density point of that clump,
$\mathbf{v_{\mathrm{cluster}}}$ is the bulk velocity of the cluster,
$\mathbf{v}_{\mathrm{clump}}$ is the bulk velocity of the clump,
$M_{\rm clump}$ is the total mass of the clump (gas + dark matter),
and $M_{\rm cluster}$ is the total mass of the cluster.
The first term enforces a preference for clumps closer to the overall center of mass,
the second attempts to eliminate clumps of gas moving quickly compared
to the bulk of the cluster (such as infalling clumps of gas),
and the third chooses larger clumps over smaller
(which has the effect of eliminating small clumps of infalling gas and is complementary to the second term).

\section{Comparison Observational Dataset: \hif}\label{sec:ObsData}

Our main comparisons against observations use the set of clusters
analyzed in \citetalias{2010A&A...513A..37H}, which includes all in the \hif\ sample
\citep[originally defined in][]{2002ApJ...567..716R}.
The \hif\ is a statistically complete, flux-limited sample of clusters with
$S_{0.5-2keV} \geq 2\times10^{-11}$ ergs/sec/cm$^2$.
This sample is comprised of 64 of the X-ray brightest clusters at high galactic latitudes.
The average redshift for the rich clusters in the sample is 0.053.
All clusters in the \hif\ sample have been observed by both Chandra and XMM-Newton with
good signal/noise, and most have been observed multiple times.
The median Chandra and XMM-Newton integration times are 65 ksec and 68 ksec, respectively.
\hif\ has a significant number of both cool core and non-cool core clusters,
mostly Abell clusters, but also includes a sample of galaxy groups, as well.
The \hif\ observations are analyzed
in \citetalias{2010A&A...513A..37H}
by a CIAO + CALDB pipeline described in detail in \citet{2006A&A...453..433H}.
The pipeline produces mosaiced X-ray images that are used to 
calculate 16 observable quantities for each cluster.

The goal of work presented in  \citetalias{2010A&A...513A..37H} is to investigate which physical properties of
clusters can be used, and with what confidence, to differentiate between types of clusters.
Using a statistical test for Gaussian bimodality (or trimodality if the algorithm gave it a higher confidence;
\citealt{1994AJ....108.2348A})
the clusters are divided into two (or three) subgroups.

The subgroups for each observable are labeled according to the
physical interpretation of the differences between subgroups.
In the case of a bimodal distribution, the clusters are divided into the usual
CC and NCC subgroups.
For trimodal distributions, the CC subgroup is divided into Strong-Cool Core (SCC)
and Weak-Cool Core (WCC) subgroups.
The cut(s) between subgroups are defined by the Gaussian test,
and clusters are assigned to subgroups based on where they fall relative to the cuts.

The main results of \citetalias{2010A&A...513A..37H} are summarized with
histograms and Gaussian fits for each quantity in their Fig. 4.
All quantities show high confidence of at least bimodality.
Two quantities, central cooling time and entropy, are further deemed to be most likely trimodal.
In order to establish the ``defining'' subgroups (those to which the Gaussian test gives the highest statistical likelihood)
between types of clusters, they compare the likelihood of all bimodal distributions
(including a {\it bimodal} distribution of central cooling time and entropy).
They find that using {\it bimodal} distributions of cooling time and entropy give the most statistically likely subgroups,
and therefore they describe these two as the defining parameters of cluster type.
As previously mentioned, they find that further dividing central cooling time and central entropy into three subgroups is actually more likely than two,
resulting in the {\it trimodal} central cooling time and entropy subgroups as the defining subgroups for cluster type.
The trimodal cuts for cooling time as defined by \citetalias{2010A&A...513A..37H} are
SCC $\leq$ 1 $h^{-1/2}$ Gyr, 1 $h^{-1/2}$ Gyr $<$ WCC $\leq$ 7.7 $h^{-1/2}$ Gyr, and 7.7 $h^{-1/2}$ Gyr $<$ NCC.
For entropy they are SCC $\leq$ 22 $h^{-1/3}$ keV cm$^2$, 22 $h^{-1/3}$ keV cm$^2$ $<$ WCC $\leq$ 150 $h^{-1/3}$ keV cm$^2$,
 and 150 $h^{-1/3}$ keV cm$^2$ $<$ NCC.
Most clusters lie in matching central cooling time and central entropy subgroups,
but 5 clusters lie just across dividing lines and have different designations (see Fig. 5 of \citetalias{2010A&A...513A..37H}).

In \citetalias{2008ApJ...675.1125B} we define a CC cluster as one with a $\geq20\%$ drop in
temperature to the center when compared to the temperature of the gas where the drop begins
(i.e. where the inward radial temperature profile slope becomes negative).
This is qualitatively similar to the definition used in \citetalias{2010A&A...513A..37H},
which we will adopt in this paper.
They define the central temperature ratio\footnote{\citetalias{2010A&A...513A..37H} uses the term
``central temperature {\it drop},'' which we modify in recognition of the many observed
clusters with $T_0/T_{\rm vir}>1$.}
as $T_0/T_{\rm vir}$,
where $T_0$ is the central temperature,
and $T_{\rm vir}$ is the ``temperature of the X-ray emitting gas that is in hydrostatic equilibrium with the cluster
potential.''
$T_{\rm vir}$ is found using a mass-scaling relation (see \S\ref{sec:CentralTemperatureDropDef} for details).
They find that a single {\it bimodal} cut between CC/NCC clusters of $T_0/T_{\rm vir}=0.7$ gives them one of
their most statistically confident bimodal distributions.

Due to the high confidence of these three parameters (central cooling time, entropy, and temperature ratio)
as defining characteristics,
and our previous use of one of them,
we will focus our observational comparisons on these quantities.

We make use of several other observational datasets for additional comparisons,
but these are not used to constrain our simulations.
We compare our simulated cluster metallicities to results presented in \citet{2011A&A...527A.134M},
and our star formation rates to values from \citet{2007ApJ...670..928B}
and \citet{2004ApJ...615..209H, 2007ApJ...654.1175H}.

\subsection{The Definition of Central Region}

The definition of the central region in \citetalias{2010A&A...513A..37H}
is the volume enclosed inside the innermost annular bin
of the radial profiles derived from the X-ray maps.
The physical size of this region depends on many things,
in particular the size of the cluster itself and the number of source counts.
We cannot use this method for our simulated data -- we can, however, infer an upper limit.
The central cooling time is calculated by \citetalias{2010A&A...513A..37H}
(see their Eqn. 15, and our \S\ref{sec:CoolingTimeDef}) using the average
temperature inside $r\leq0.048$ $R_{500}$,
which implies that the central region is no larger than this.
For some of the smaller simulated clusters,
using this radius for the central region would result in calculating quantities
over just a few cells.
To avoid stochasticity due to the inclusion of very small numbers of cells, we define the central region as the
inner $r\leq50$ kpc for all clusters, which covers $\sim50$ maximum-resolution cells.
This radius was chosen after investigating the values
of our observables over a range of central region radii.
Using a radius less than 50 kpc gives a scatter of values that
covers many orders of magnitude, and is much larger scatter than observations.
Choosing a radius larger than 50 kpc showed very little change in cooling time beyond the 50 kpc values;
no larger than a factor of two out to several hundred kpc.
50 kpc was therefore chosen as a compromise between volumes too small
and volumes much larger than used in the observational data analyses.
In fact, for the majority of the clusters, 50 kpc is within a factor of two of
$0.048$ $R_{500}$, and is therefore a reasonable estimate for the central radius.

\subsection{Our Three Main Observables}\label{sec:SynthObs}

In the previous section, we identified the central cooling time, entropy, and temperature ratio
as the three quantities upon which we will focus our comparisons.
In this section, we describe how we calculate each observable quantity.

\subsubsection{Spectroscopic-Like Temperature}\label{sec:TspecDef}

In place of the simulated gas temperature $T$,
we use the spectroscopic-like temperature $T_{\textrm{spec}}$,
calculated as follows:

\begin{equation}
T_{\textrm{spec}} \equiv \frac{ \int \rho^2 T^{(\alpha-0.5)} dV }{ \int \rho^2 T^{(\alpha - 1.5)} dV},
\end{equation}
over the volume of interest, where $\alpha=0.75$.
Previous work has shown that spectroscopic-like temperature
\citep[see][]{2004MNRAS.354...10M, 2005ApJ...618L...1R} is a reasonable proxy for the
measured X-ray spectral temperature.
When making mock temperature maps, we perform a similar calculation,
but use this weighting for the line-of-sight integral, as has been used in earlier studies
\citep[e.g.,][]{2010ApJ...725.1053H}.
This weighting has been shown to reproduce
the fitted spectral temperature better than standard emission weighting
or mass weighting of the temperature in simulations
\citep{2004MNRAS.354...10M}. 

We have discovered in prior work that
$T_{spec}$ has some critical limitations in the context of simulations
with radiative cooling. $T_{spec}$ in its original incarnation is a
calibrated weighting, using simulations of clusters with mean
temperatures greater than $T = 2keV$. Additionally, when calculating
$T_{spec}$, \citet{2004MNRAS.354...10M} ignored gas particles with
$T<0.5keV$. In regions in some of our simulated clusters where cooling
dominates, the gas reaches much lower temperatures than this. Because
gas at these low temperatures contributes negligibly to the X-ray
emission, including these zones in the $T_{spec}$ calculation results
in temperatures that are not representative of the X-ray
temperature for these clusters.
Therefore, in order to more accurately model the X-ray temperature for the clusters,
the cold gas should be removed from the calculation.
We discuss the implications of this and our choice to include or not include the cold
gas in clusters in \S\ref{sec:coldgas}.

\subsubsection{Central Cooling Time}\label{sec:CoolingTimeDef}

We calculate central cooling time $t_{0,\mathrm{cool}}$ over the central region
$V_0$ as follows (adapted from Eqn. 15 of \citetalias{2010A&A...513A..37H}):

\begin{equation}\label{eqn:CoolingTime}
t_{0,\mathrm{cool}} = \frac{3}{2} \zeta \frac{(n_e'+n_i') \int_{V_0} kT_{\mathrm{spec}}dV}{n_e'^2 \int_{V_0}  \Lambda(T_{\mathrm{spec}})dV}
\end{equation}
where $n'_{e}=\sqrt{\left<n_e^2\right>}$ is the number density of electrons (a volume-weighted average),
$n'_i=n'_e/\zeta$ is the number density of ions,
$\zeta$ is the ratio of electrons to ions and is set to 1.2 following \citetalias{2010A&A...513A..37H},
and $\Lambda(T_{\mathrm{spec}})$ is the cooling function.
Values of $\Lambda$ come from the same cooling function table used in the RS simulations,
which is very similar to the \it APEC\rm\ \citep{2001ApJ...556L..91S} cooling function used in \citetalias{2010A&A...513A..37H}
for optically thin plasmas.
Using a metallicity-independent cooling function for the cooling time calculation is an extra
approximation for the simulations that use \cloudy\ cooling,
but given that most of the free-free emission at high temperatures comes from primordial gas,
this is a reasonable approximation.
We use $n_e'$, instead of the average of the simulated electron number
density $n_e$,  since X-ray observations directly measure the value of
$\langle n_e^2 \rangle$ from the thermal emission in the cluster, not
$\langle n_e \rangle$ \citep[see, e.g.,][]{2011Sci...331.1576S}.  This measurement therefore produces a bias in the X-ray inferred electron density equal to the clumping factor, C, of the gas, where 

\begin{equation}
C = \frac{\langle n_e^2 \rangle}{\langle n_e \rangle ^2},
\end{equation}
where the averages are volume-weighted.

\subsubsection{Central Entropy}\label{sec:CentralEntropyDef}

For the purpose of comparing the entropy in the simulated clusters to
the observationally deduced entropy as in the \citetalias{2010A&A...513A..37H} sample, we
calculate a slightly modified version of the entropy from our
simulation grid. While for the most part, the variation from the
standard thermal entropy are small using this method, it is not
identical. The standard thermal central entropy calculation for the simulation
thermal properties is

\begin{equation}
K_0 =  \frac{\langle kT \rangle}{n_e^{2/3}},
\end{equation}
where $T$ is the thermal temperature of each simulation zone in the
central region, and $n_e$ is the mass-weighted average of the central
electron density in those same regions. When comparing to
observational data, however, this is not precisely analogous. The
deduced temperature from X-ray data is the spectroscopic temperature,
which is not exactly equal to the mass-weighted mean gas temperature in
our simulations.
Therefore, we replace the average of $kT$ in this
calculation with the value of $kT_{\textrm{spec}}$, which is the
spectroscopic-like temperature of \citet{2005ApJ...618L...1R}, described in \S\ref{sec:TspecDef}.
We also use $n_e'$ (defined in \S\ref{sec:CoolingTimeDef}) in the
entropy calculation to account for the bias introduced by X-ray
determinations of the electron density, and allow a clean comparison
with observations. A more observationally analogous quantity to be
generated from our simulations is then

\begin{equation}\label{Eqn:Entropy}
K_0 = \frac{kT_{\mathrm{spec}}}{n'^{2/3}_e}
\end{equation} 
over the central region of interest, which we use for our comparisons to observations. 

\subsubsection{Central Temperature Ratio}\label{sec:CentralTemperatureDropDef}

We calculate the central temperature ratio as
the ratio of the central temperature $T_0=T_{\mathrm{spec,}0}$
over the virial temperature $T_{\mathrm{vir}}$.
The virial temperature is calculated from a scaling relation of $M_{500}$ ($500\times$ the mean background density;
Eqn. 4 of \citetalias{2010A&A...513A..37H}):

\begin{equation}\label{eqn:Tdrop}
T_{\textrm{vir}} = \frac{1\mathrm{ keV}}{k} \log_{1.676}\left[\frac{M_{500}}{2.5\times10^{13} h^{-1} M_\odot}\right].
\end{equation}
We use this mass scaling relation instead of calculating it from the mass or spectroscopic-like weighted temperature
because it eliminates the need to remove cold gas (see \S\ref{sec:coldgas}) in some of the clusters 
(which is equivalent to determining a cluster has a cool core before we try to categorize it later).
It is appropriate for clusters at low redshift with masses similar to
our simulated clusters,
and it is also how observationally-derived masses are calculated in \citetalias{2010A&A...513A..37H}.

\section{Understanding the Effects of Varied Numerical Methods}\label{sec:FirstFourSims}

\begin{figure*}[htbp] 
   \centering
   \includegraphics[width=1.\textwidth]{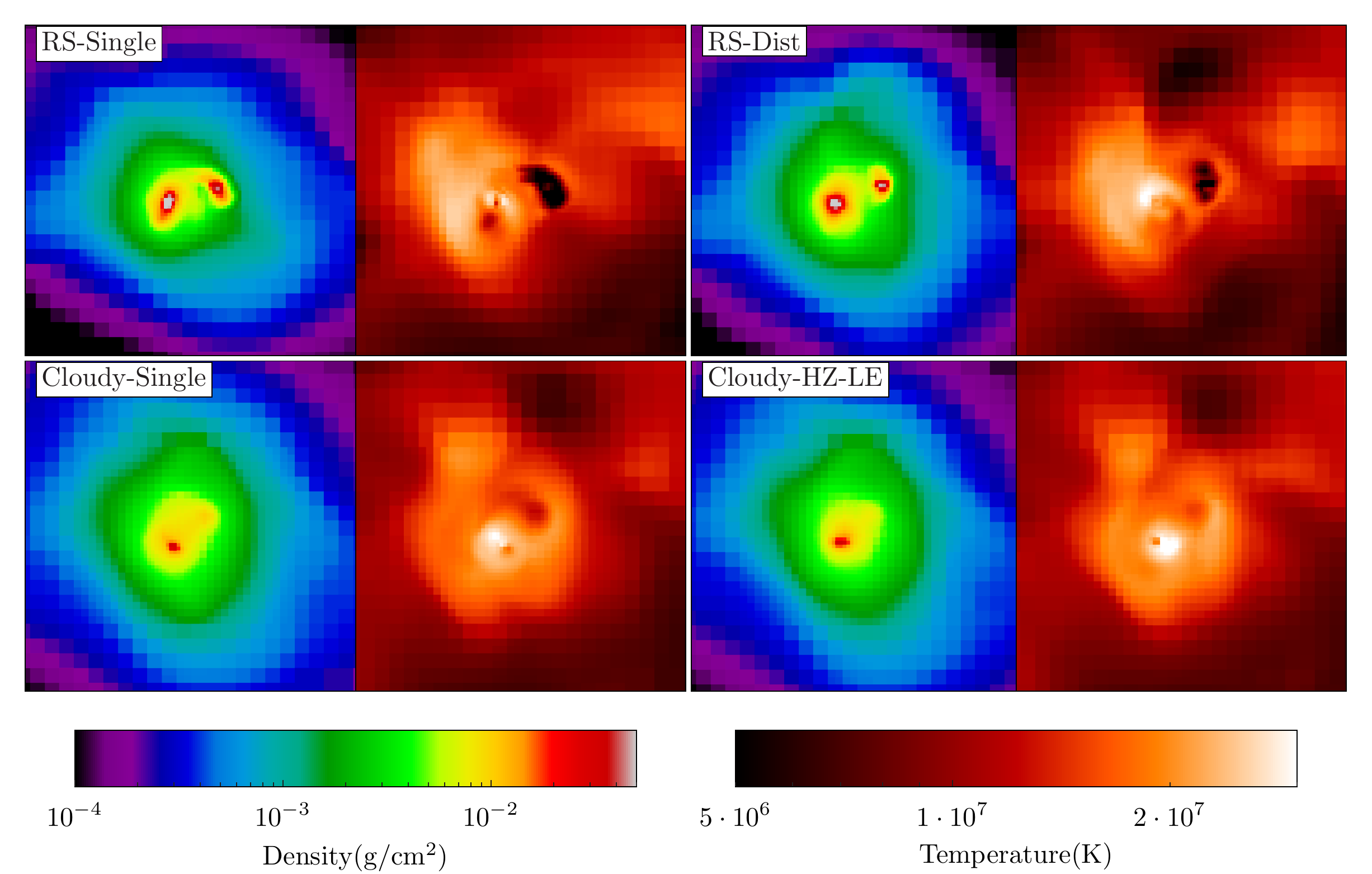}
   \caption{Projections of gas column density (left panel in each pair) and gas density-weighted temperature (right panel) of
   the corresponding halo from the four simulations as labeled.
   The density and temperature color bars apply to all four pairs.
   ``RS-Single'' and ``RS-Dist'' employ metallicity-independent gas cooling,
   while ``Cloudy-Single'' and ``Cloudy-HZ-LE'' use metallicity-dependent cooling.
   Star particles in ``RS-Single'' and ``Cloudy-Single'' deposit all their feedback into the single cell that surrounds them,
   while stars in  ``RS-Dist'' and ``Cloudy-HZ-LE'' use distributed feedback and deposit their feedback over the 27 cells that surround them.
   The projections are through a cube 2 Mpc on a side.
   The cluster has R$_{200}\approx1.9$ Mpc, M$_{200}\approx 3.9\times10^{14}$ M$_{\odot}$ and $T_{\rm vir}\approx 3.7\times10^7$ K (using Eqn. \ref{eqn:Tdrop}).}
   \label{fig:halo29}
\end{figure*}

Our first suite of simulations focus on the effects of \cloudy\ cooling (i.e., metal-dependent) and distributed feedback
on simulations and observables.
This is accomplished with four simulations beginning with an initial simulation
similar to what was performed in \citetalias{2008ApJ...675.1125B}.
The only functional difference is a somewhat higher energy feedback parameter $\epsilon_E$:
$4.11\times10^{-6}$ in \citetalias{2008ApJ...675.1125B} versus $1.0\times10^{-5}$ here.
Our current value releases an amount of energy per unit stellar mass within a factor of two
of several other similar studies of clusters \citep[e.g.][]{2005ApJ...625..588K,
2007MNRAS.382.1050T}.
A further difference is that in \citetalias{2008ApJ...675.1125B} each cluster is simulated in a
separate zoom-in simulation,
while in this work we simulate all of the clusters at once.
This should not, however, lead to any meaningful differences because the mass and peak spatial resolutions,
refinement criteria, and the physics modules are identical in the two cases.

We label the first simulation ``RS-Single'' (i.e. metal-independent cooling with feedback deposited in a single cell)
 to make clear which cooling and feedback modules are in use.
The remaining three simulations (see Table \ref{table:SimParams})
swap RS cooling in favor of \cloudy\ cooling (``Cloudy-Single''),
swap single cell for distributed feedback (``RS-Dist''),
or swap both (``Cloudy-HZ-LE'', which can be read as ``Cloudy-High Metal Feedback-Low Energy Feedback'').
We do not change anything else between simulations,
and in particular, $\epsilon_E$ and $\epsilon_Z$ are kept constant between simulations.
The figures in this section include all the gas in the central regions of the clusters,
which may include cold gas invisible to X-ray observatories.
We discuss the impact of the cold gas on our calculations in \S\ref{sec:coldgas}.

\begin{figure}[!h] 
   \centering
   \includegraphics[width=0.45\textwidth]{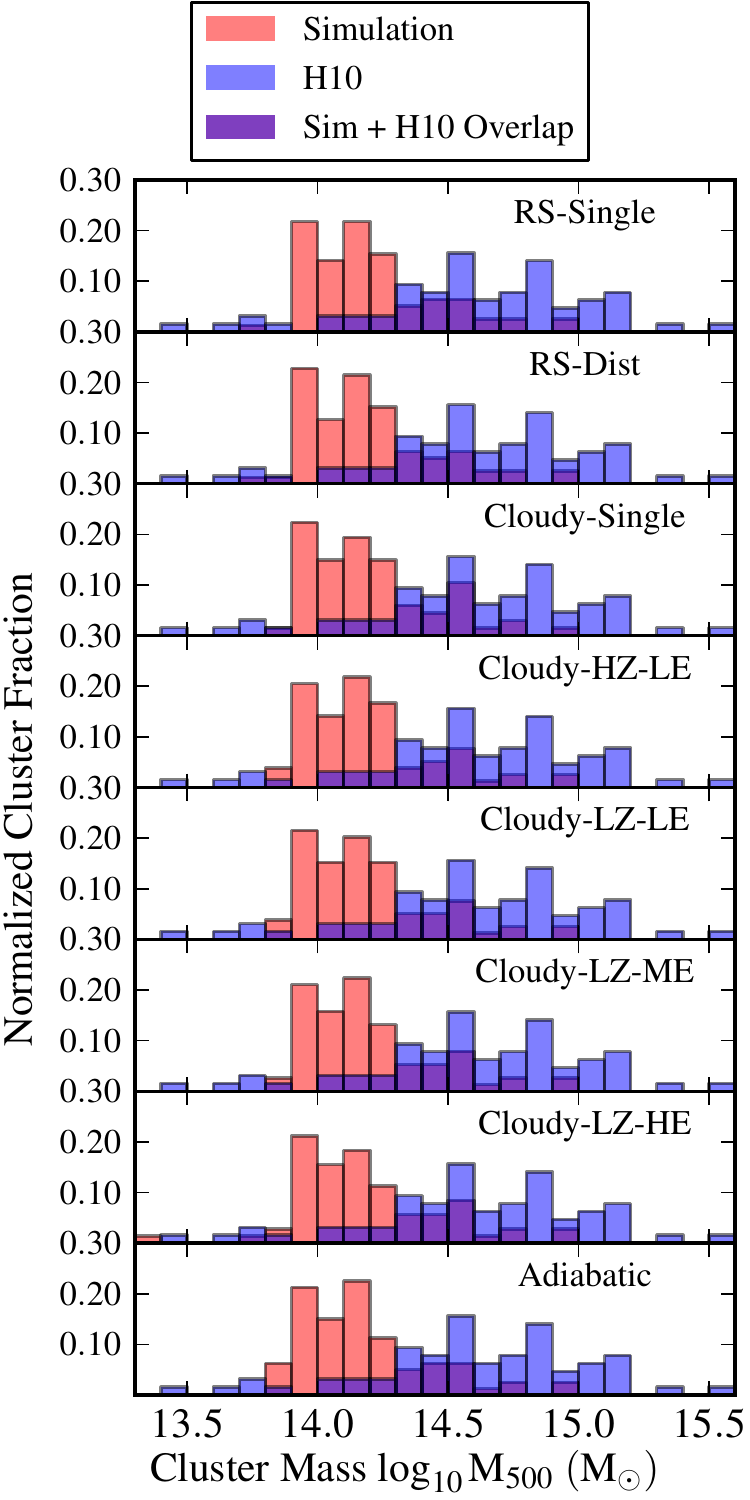}
   \caption{Normalized distribution of M$_{500}$ for simulated and observed clusters.
   The colors are semi-transparent -- dark blue indicates overlap between observation and simulation.
   The values of M$_{500}$ for \citetalias{2010A&A...513A..37H} data are found using their Eqn. 4
   and $kT_{\mathrm{vir}}$ from their Table 2.
   The simulations in the lower four panels are introduced and discussed in Sections
   \ref{sec:AdjustFeedback} and \ref{sec:Discussion}.}
   \label{fig:histM500}
\end{figure}

Additionally, below we compare the observable quantities of our simulated clusters against the observations
presented in \citetalias{2010A&A...513A..37H}.
Figure \ref{fig:histM500} shows that our most massive clusters are less massive than the most massive observed clusters.
Therefore, for purposes of comparing simulations to observations,
a subset of the \citetalias{2010A&A...513A..37H} sample has been selected to exclude clusters
with M$_{500}$ values greater than the most massive simulated clusters.
However, as we will discuss in \S\ref{sec:HighMassClusters}, more massive simulated clusters
do not exhibit substantially different behavior than the clusters discussed in this section.
This selection eliminates 3 of 28 SCC clusters, 4 of 18 WCCs, and 5 of 18 NCCs from the observational sample,
but does not greatly affect the overall ranges, nor distribution of observed
central cooling time, entropy, or temperature ratio.

\subsection{The Impact of Metallicity-Dependent Cooling}\label{sec:ABCloudy}

\begin{figure}[htbp] 
   \centering
   \includegraphics[width=0.45\textwidth]{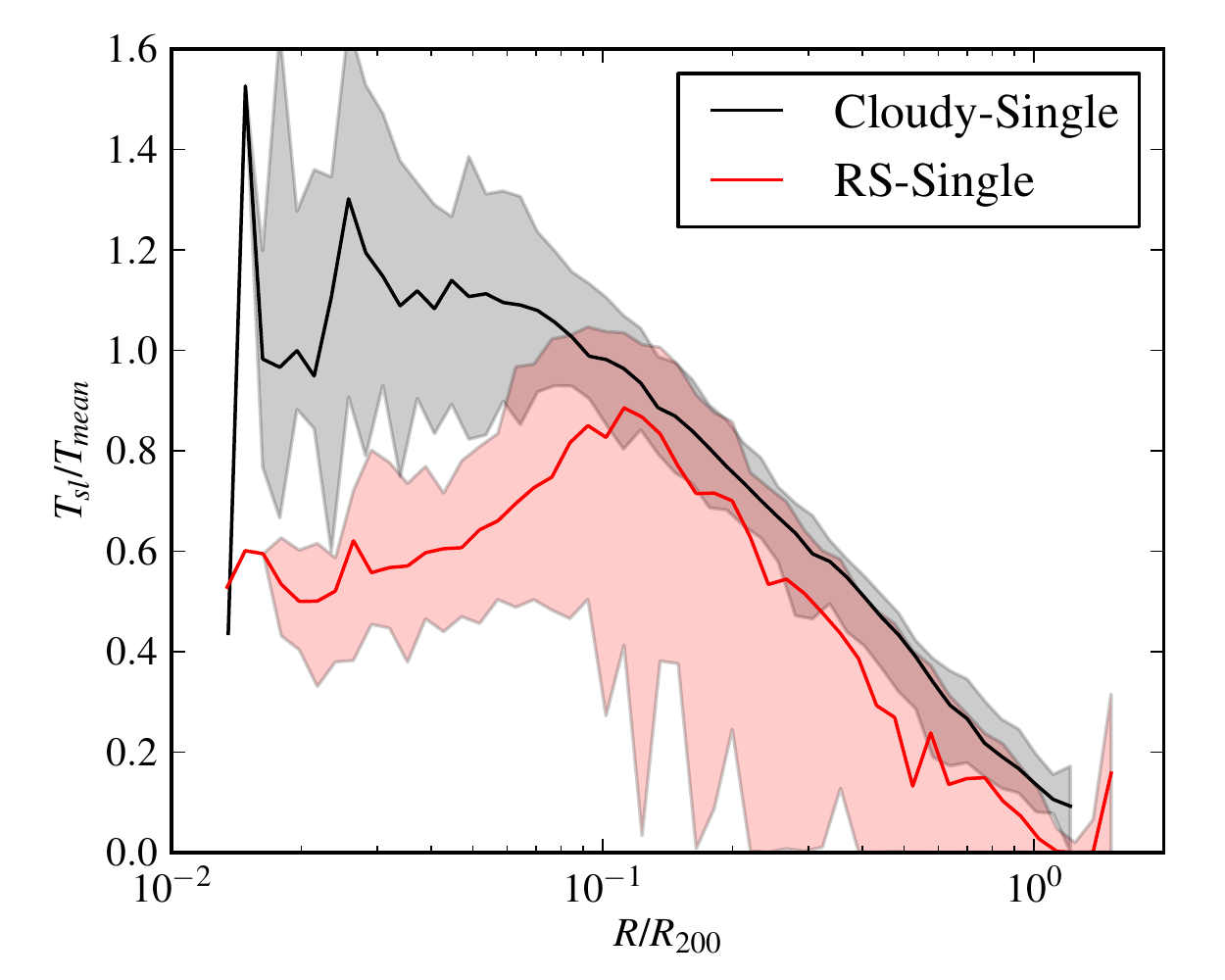}
   \caption{The mean (solid lines) normalized temperature profiles for all
   Cloudy-Single (blue) and RS-Single (red) clusters
   and the $1\sigma$ scatter (semi-transparent filled regions).
  }
     \label{fig:TprofCC_RC}
\end{figure}

The left column of panels in Figure \ref{fig:halo29} visually compares RS-Single with Cloudy-Single
using a representative cluster.
This particular cluster is used because it is undergoing a single merger,
which allows us to illustrate several important differences between simulations,
but it is not overly complicated by multiple simultaneous mergers.
It is clear that the choice of cooling method has a pronounced effect on the cluster.
The Cloudy-Single cluster is less centrally dense than the same cluster in the other three models,
and the infalling clump that is very cold in
the RS-Single case is much warmer with smoother gradients in, e.g., Cloudy-Single.
These differences are also evident in the temperature profiles shown in Fig. \ref{fig:TprofCC_RC}.
At all radii, the mean Cloudy-Single cluster is warmer than the mean RS-Single cluster.
Outside $0.1R_{500}$, the profiles are shifted by a nearly constant amount ($\sim 0.1$),
which is a direct result of \cloudy\ cooling preventing unenriched gas at high redshift from becoming too cold.
The large differences in temperatures inside $0.1R_{500}$ are also a result of
the early overcooling of the RS method due to its assumption
of relatively high metallicity at all times.

\begin{figure*}[htbp] 
   \centering
   \includegraphics[width=1.0\textwidth]{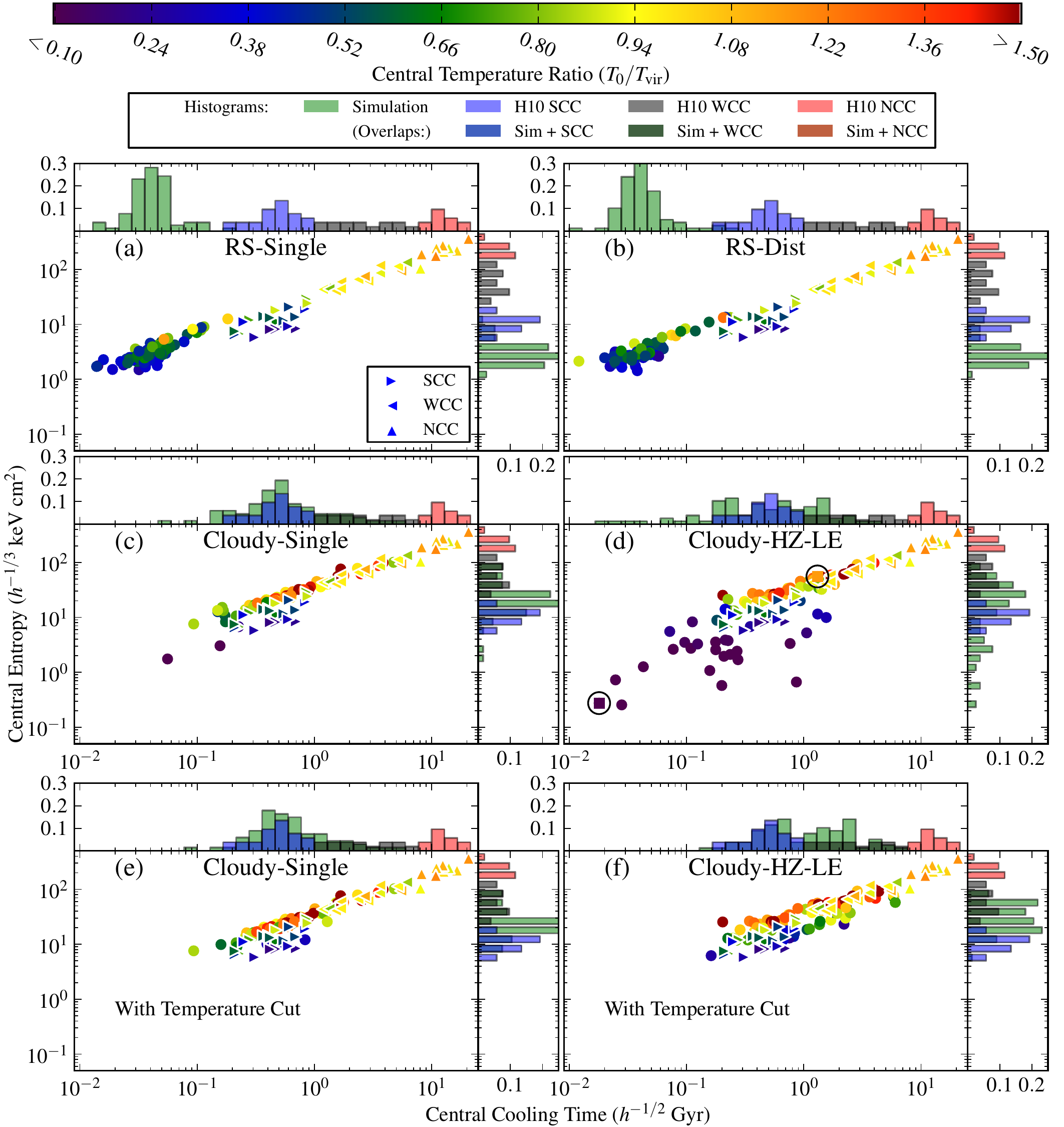}
   \caption{Comparison of central cooling time, entropy, and temperature ratio for both simulated and observed clusters.
    Please see text for a detailed description of this figure.
    The circled points in panel (d) correspond to the two clusters in Fig. \ref{fig:CD_mergers}.
   }
   \label{fig:cc-cuts2}
\end{figure*}
In Fig. \ref{fig:cc-cuts2} we compare simulations against observations,
and by extension, to one other.
Figure \ref{fig:cc-cuts2} contains a great deal of information and therefore warrants a detailed description.
For a given simulation, there are three elements: the main panel and two histograms.
The main panel (which has the largest area and contains the simulation label) shows the central cooling time
versus entropy for both simulations and observations, with the simulated clusters represented by
circles and the sub-selected \citetalias{2010A&A...513A..37H} cluster sample represented by triangles.
The \citetalias{2010A&A...513A..37H} triangles are oriented according to the central cooling time subgroup cuts
(SCC $\leq1$ h$^{-1/2}$ Gyr, 1 $<$ WCC $\leq7.7$, and 7.7 $<$ NCC).
The colors of the glyphs in the main panel correspond to the central temperature ratio
following the color scale along the top of the figure.
Both above and to the right of each main panel are normalized histograms
of central cooling time (above) and entropy (right) for simulations and \citetalias{2010A&A...513A..37H} data.
The colors of the histogram bars indicate the type of data plotted
(simulated or \citetalias{2010A&A...513A..37H} subgroups) following the histogram
color key, which is above all the figure panels and below the color bar.
The entropy histograms are cut according to
SCC $\leq$ 22 $h^{-1/3}\ \mathrm{keV}\ \mathrm{cm}^2$, 22 $<$ WCC $\leq$ 150, and 150 $<$ NCC.
Histogram bars are semi-transparent, and overlap between simulation and
\citetalias{2010A&A...513A..37H} data are indicated by the three darker colors
in the histogram color key.
Note that the entropy distributions for panels (a),(b), and (e) are highly peaked
and one or more histogram bar extends beyond the displayed fractional range (0.25).

\begin{figure}[htbp] 
   \centering
   \includegraphics[width=0.45\textwidth]{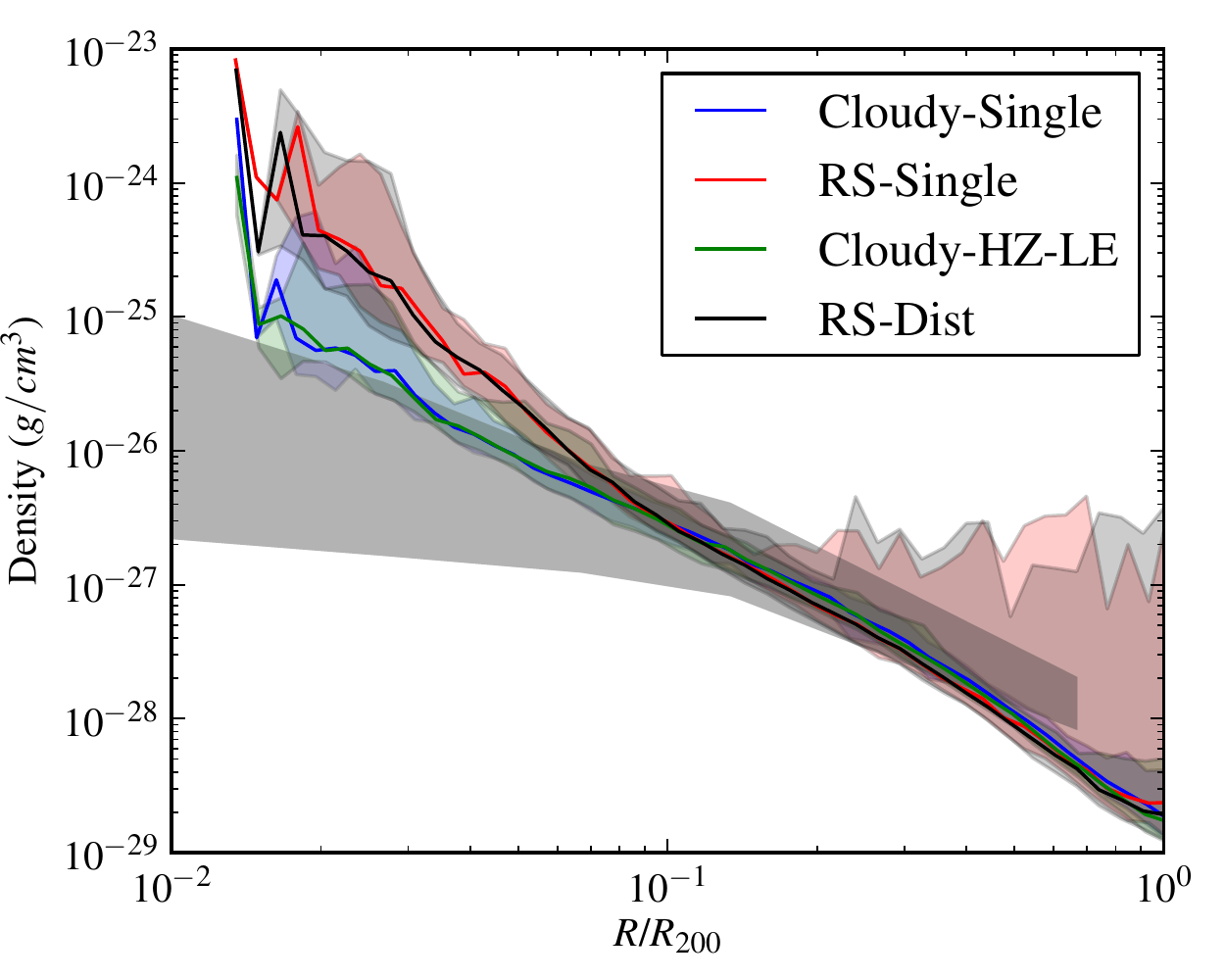}
   \caption{Mean density profiles for each simulation (solid lines) with $1\sigma$ scatter
   (semi-transparent filled regions). The range of observed densities from \citet{2011A&A...536A...9P}
   is over-plotted in grey.}
   \label{fig:All4DensityProfiles}
\end{figure}

We can now remark on the profound effect \cloudy\ cooling
has on the observables between panels (a) and (c) of Fig. \ref{fig:cc-cuts2}.
While central cooling times are much shorter than observed in the RS-Single case,
Cloudy-Single clusters exhibit a much wider distribution of cooling times
that are shifted to higher, more physically realistic values.
A similar effect is seen in the central entropy values as well.
However, despite the apparent visual improvement, non-parametric Kolmogorov-Smirnov (KS) \citep{1933Kolmogorov, 1939Smirnov}
tests between data and simulation for these two observables gives $p$-values less than 0.01 (the typical lower cutoff for statistical similarity)
for both RS-Single and Cloudy-Single,
which indicates that none of the simulated distributions are statistically similar to observations.
The short cooling times and low entropies with RS cooling
are a direct result of early universal overcooling.
The overall colder gas produces higher central cluster densities
(Fig. \ref{fig:All4DensityProfiles}),
and lower central temperatures (e.g., Fig. \ref{fig:TprofCC_RC}) which, in the relevant
temperature range ($\sim 10^7$ K), result in higher cooling rates (Eqn. \ref{eqn:CoolingTime}).
Note that Fig. \ref{fig:All4DensityProfiles} shows that outside the core, both methods of cooling
produce cluster density profiles in rough agreement with observations,
but that towards the core all simulations show central densities higher than observations,
which is a typical problem with these kinds of simulations.
The use of \cloudy\ cooling creates clusters with less
unphysically high central densities than with RS cooling,
but additional physical models (e.g. AGN feedback) may be required to address this discrepancy fully.

\begin{figure}[!h] 
   \centering
   \includegraphics[width=0.45\textwidth]{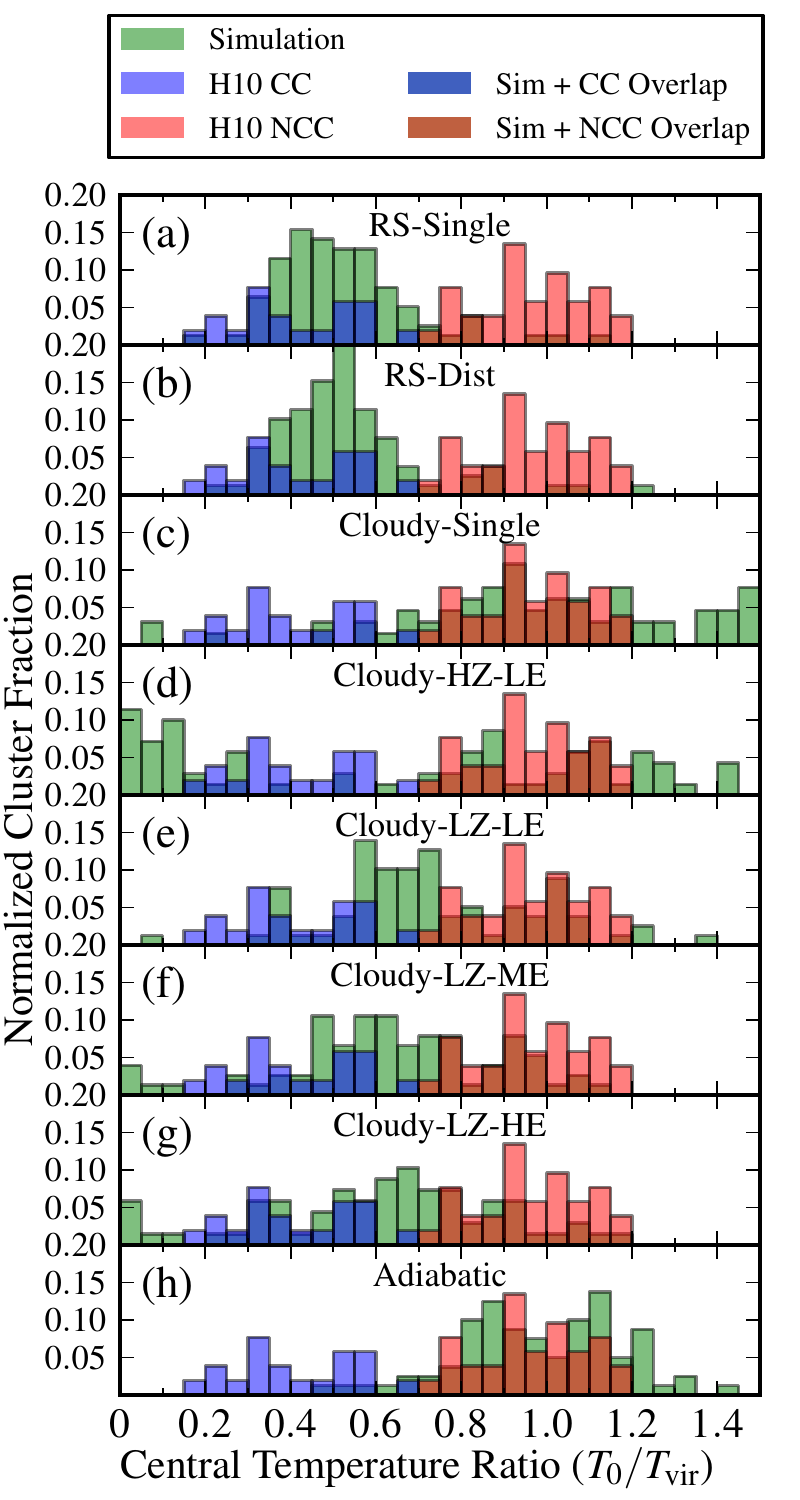}
   \caption{Central temperature ratio histograms for simulated clusters (green)
   and observed clusters of comparable mass.
   The calculations for the simulated clusters do not include a temperature cut.
   The colors for observed data correspond to the
   bimodal groups NCC (red) and
   CC (blue) as defined by \citetalias{2010A&A...513A..37H}
   (CC $\leq$ 0.7 = $T_0 / T_{\mathrm{vir}}$).
   The simulations in the lower four panels are introduced and discussed in Sections
   \ref{sec:AdjustFeedback} and \ref{sec:Discussion}.}
   \label{fig:hist-Tdrop}
\end{figure}

In Figure \ref{fig:hist-Tdrop} we compare the central temperature ratios of simulated
and observed clusters.
It's clear that neither RS-Single (a) nor Cloudy-Single (c) are similar to observations.
RS-Single is too cool and produces few NCC clusters, while Cloudy-Single is the opposite,
with too many NCC clusters.
Neither of the KS tests between these two simulations and the observed distribution of central temperature ratios
give $p$-values greater than 0.01.

\subsection{The Impact of Distributed Supernova Feedback}

\begin{figure}[htbp] 
   \centering
   \includegraphics[width=0.45\textwidth]{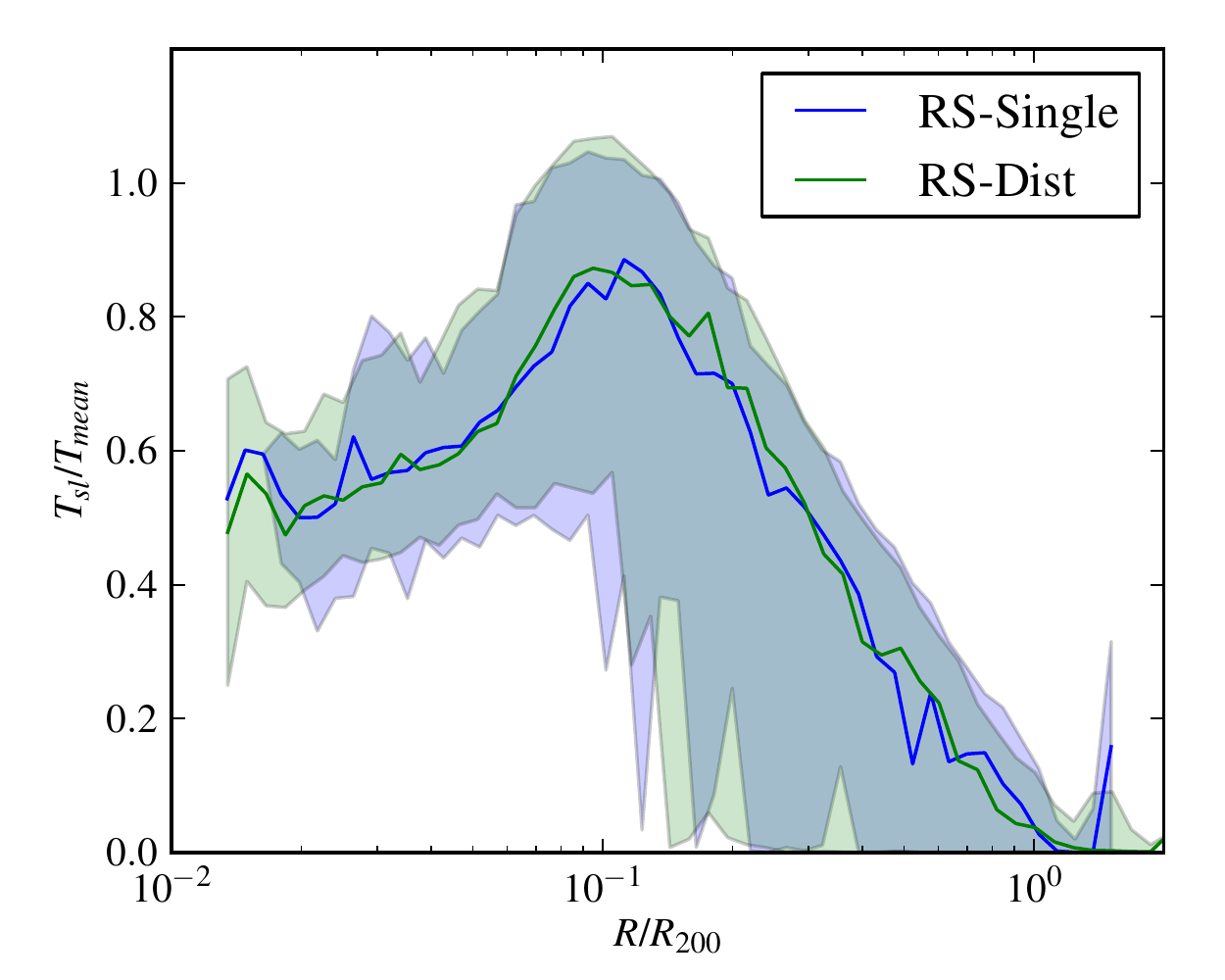}
   \caption{Similar to Fig. \ref{fig:TprofCC_RC},
   except the simulations used are RS-Single (blue line) and RS-Dist (green line).
   }
      \label{fig:TprofRC_RD}
\end{figure}

Next, we study the effect of varying the spatial extent of thermal and metal feedback from supernovae
(using RS cooling in both simulations).
Figure \ref{fig:TprofRC_RD} shows that the mean temperature
profiles of RS-Single and RS-Dist are indistinguishable.
This is a reasonable result because the addition of distributed feedback
does not change the total amount of energy deposited into the gas by a given star particle.
Interestingly, there are some differences that can be seen when the temperature
projections are compared in Fig. \ref{fig:halo29}.
From RS-Single to RS-Dist, it appears that the amount of very cold gas in the
infalling clump is somewhat diminished.
This is consistent with the distributed energy feedback of merger-driven star formation
more effectively heating the cold gas of the clump over a wider region.
This same effect is seen in projections of other merging cold clumps in RS-Dist clusters.
Note that although although the infalling clump has less cold gas with distributed feedback,
the mass of the clump is low enough (and other merging clumps like it)
that averaged over all the clusters, the warming effect is not enough to dramatically change
the mean temperature profile between the two simulations.

The effect of distributed feedback on observables is smaller than the changes when
\cloudy\ cooling is used, which is illustrated in Fig. \ref{fig:cc-cuts2}.
The distributions of central cooling times and entropies of the RS-Dist clusters are slightly widened,
but do not resemble observations.
Likewise, the spread of temperature ratios (Fig. \ref{fig:hist-Tdrop}) is slightly greater with distributed feedback,
but again is not very similar to observations.
Like RS-Single, none of the KS tests for RS-Dist indicate that the simulated distributions are statistically similar to observations.
Considering all three observables together, 
it appears that distributed feedback has a slightly similar, but much less strong effect on the observables as does \cloudy.
Namely, by distributing feedback over wider regions, including cold, tight clumps (like in Fig. \ref{fig:halo29})
it acts to partially counter the RS overcooling and widen the distributions of the observables.
The entropies for both RS simulations are low because nearly all the clusters have cool central temperatures
and
central densities higher than \cloudy\ simulations (Fig. \ref{fig:All4DensityProfiles}).
These two effects are both a consequence of early overcooling,
and are complementary in Eqn. \ref{Eqn:Entropy},
resulting in low entropies.

\subsection{The Impact of Combining Distributed Feedback + Metallicity-Dependent Cooling}

Finally, we examine the result when both distributed feedback and \cloudy\ cooling are used simultaneously.
Morphologically, the combination of \cloudy\ cooling and distributed feedback appears
to only make small differences when compared to the substitution of only \cloudy.
Indeed, the qualitative differences between rows of Fig. \ref{fig:halo29} are greater
than the differences between columns.
{\it This highlights the importance of the addition of metallicity-dependent cooling due to its
greater effect on the clusters when compared to the addition of distributed feedback.}
This is reasonable because distributed feedback
is only operational when and where star particles are being formed,
while \cloudy\ cooling makes substantial changes to cooling rates compared to RS cooling over the whole
simulation at all times.

\begin{figure}[htbp] 
   \centering
   \includegraphics[width=0.45\textwidth]{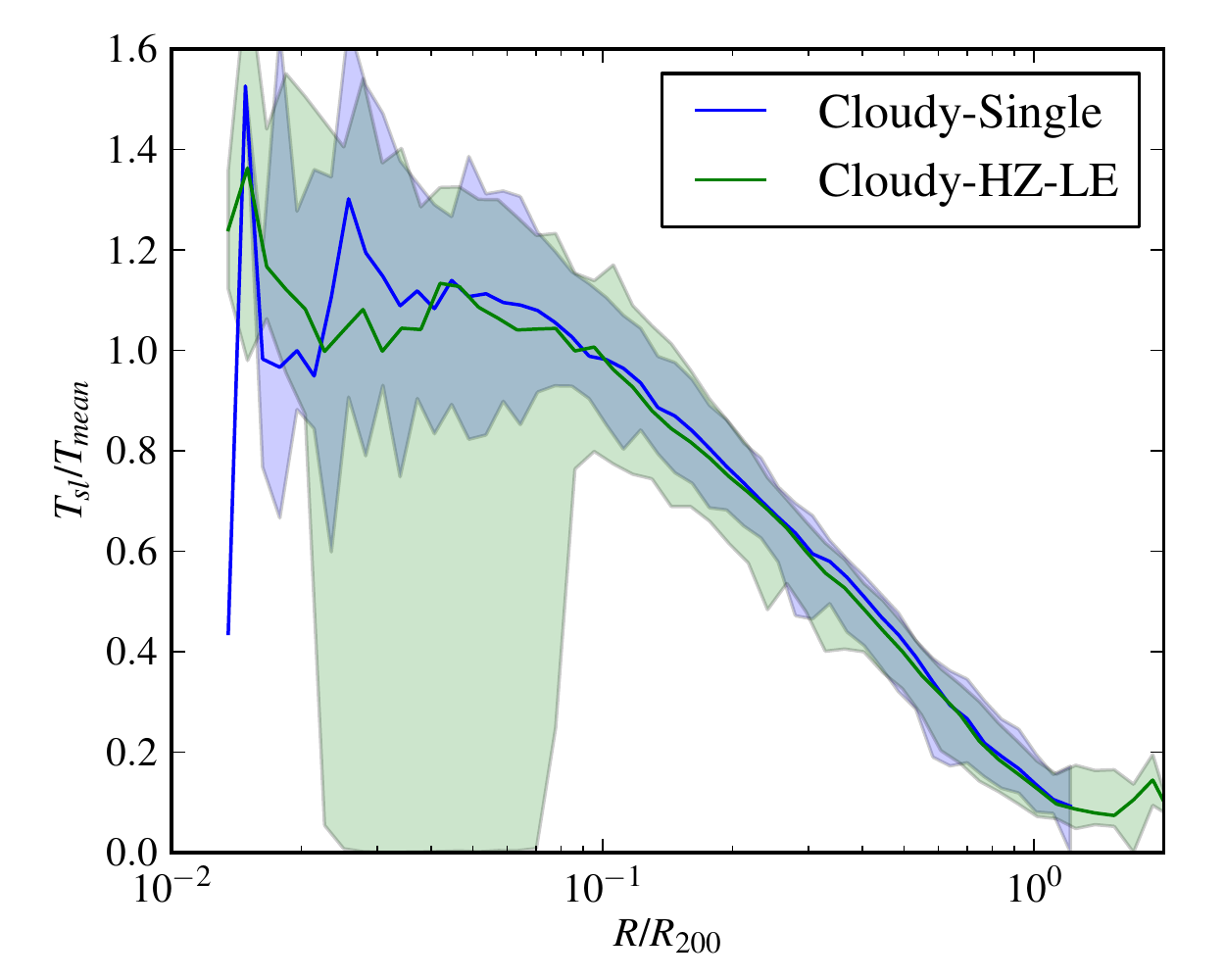}
   \caption{Similar to Fig. \ref{fig:TprofCC_RC},
   except the simulations used are Cloudy-Single (blue line) and Cloudy-HZ-LE (green line).
   }
      \label{fig:TprofCC_CD}
\end{figure}

Figure \ref{fig:TprofCC_CD} illustrates this point.
Outside $0.1R_{500}$, which accounts for the bulk of the cluster,
the temperature profiles are nearly identical.
However, inside $0.1R_{500}$, where the star particles are applying feedback,
the differences between Cloudy-Single and Cloudy-HZ-LE become much larger.
In particular, the $1\sigma$ scatter of Cloudy-HZ-LE temperatures is much larger than it is for Cloudy-Single.
Although Cloudy-Single produces approximately 25\% more star particles by mass than Cloudy-HZ-LE
(see Fig. \ref{fig:SFR} in \S\ref{sec:LackRes}),
the ratio of total mass of metals (in star particles and in the gas) to the total mass of stars is the same within few percent
in both simulations.
The difference is that in Cloudy-HZ-LE, 19\% of the metals by mass are stored in the gas,
while this figure is only 10\% for Cloudy-Single,
which implies that the Cloudy-HZ-LE has $\sim50$\% more metals by mass in the gas
than Cloudy-Single.
The increased metals in the gas allows for much higher rates of cooling,
giving rise to the scatter seen in Fig. \ref{fig:TprofCC_CD}.

The fact that more metals are stored in star particles in Cloudy-Single is very important.
Although some of the metals given to star particles are recycled and enriched
back into the gas after the particle is formed (see Eqn. \ref{eqn:Z}),
the metals that remain in star particles after the feedback is applied (e.g., after $12 t_{\mathrm{dyn}}$) are locked in forever.
Crucially, these metals are no longer available to contribute to cooling of the intracluster medium.
The net effect is that, in comparison to the Cloudy-HZ-LE star particles,
the Cloudy-Single star particles behave as a kind of ``metal sink.''

This result of enhanced metal feedback in the gas is a curious secondary effect of distributed feedback.
It appears that without distributed feedback, the metals returned by star particles in each star-forming region stay
contained within that region, and are re-captured by the next generation of star particles.
Distributed feedback appears to disrupt this cycle
because it deposits much of the metal-enriched gas outside star formation regions,
preventing metals from being locked in star particles,
and resulting in more metal mass in the intracluster medium.
We compare the metallicities for these two simulations to observations in \S\ref{sec:MetalComp}.

The impact of the higher metallicity and cooling in Cloudy-HZ-LE is starkly visible in Fig. \ref{fig:cc-cuts2}(c,d).
Instead of a tight linear grouping like the Cloudy-Single clusters,
the Cloudy-HZ-LE clusters are scattered widely in central cooling time, entropy, and temperature ratio.
Many clusters exhibit quite low cooling times, entropies, and extremely low central temperatures,
and some of these ``cold-core'' clusters have $T_0 / T_{\mathrm{vir}}$ much less than 0.01 and have
central temperatures that drop below $10^6$ K.
The cold-cores are tightly bunched at the low end of the distribution in Fig. \ref{fig:hist-Tdrop}(d).
It is these cold-cores that give rise to the scatter in Cloudy-HZ-LE cluster temperature profiles inside 0.1 $R_{200}$ (Fig. \ref{fig:TprofCC_CD}).
As with the earlier simulations, none of the Cloudy-HZ-LE distributions are statistically similar to observations
according to the KS test.
We note that in reality, X-ray observations cannot detect the cold gas present in the cold-cores,
which we discuss in more detail in \S\ref{sec:coldgas}.

\begin{figure}[htbp] 
   \centering
   \includegraphics[width=0.45\textwidth]{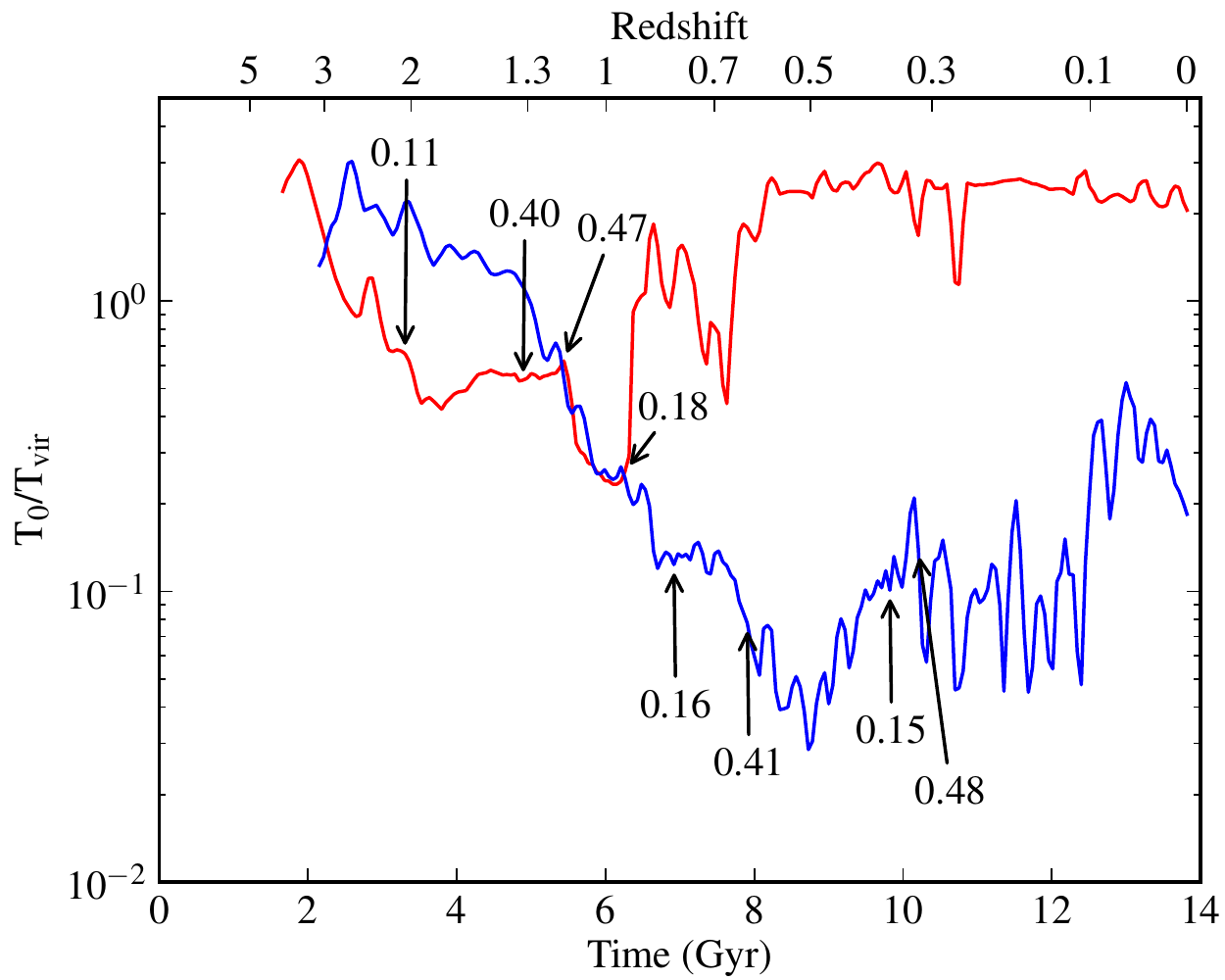}
   \caption{T$_0$/T$_{\mathrm{vir}}$ as a function of time for the two Cloudy-HZ-LE clusters indicated by
   circles in Fig. \ref{fig:cc-cuts2}(d).
   The red line corresponds to the warm cluster in the upper right,
   and the blue to the cold cluster in the lower left.
   The downward-pointing arrows show major
   (M$_{\mathrm{subhalo}}$/M$_{\mathrm{cluster}}$ $\geq$ 0.1)
   merger events for the warm cluster.
   Likewise, the upward arrows are for the cold cluster.
   The number with each arrow is the M$_{\mathrm{subhalo}}$/M$_{\mathrm{cluster}}$ ratio
   for that merger.
   At z=0 the warm cluster has M$_{200}\approx 6.0\times10^{14}$ M$_\odot$, and the cold cluster
   has M$_{200} \approx 1.5\times10^{14}$ M$_\odot$.}
   \label{fig:CD_mergers}
\end{figure}

Using a halo merger tree we examine the evolution of the clusters,
and, in particular, focus on the central temperature histories of the cold-core clusters.
As we found in \citetalias{2008ApJ...675.1125B}, after formation
all of the clusters in our calculations eventually become CC clusters with low central temperatures.
Later on, some of the clusters experience major mergers, which may disrupt
the embryonic cool core, and raises the central temperatures permanently.
However, some clusters grow by only minor mergers and smooth accretion early on, and the
low central temperatures can be preserved to z=0,
although they may encounter large impact parameters and/or minor mergers at late times.
These later mergers are insufficient to destroy the large and tightly-bound cool core, and only act to add more cool gas to the core.'
With the addition of \cloudy\ cooling, and the enhanced distribution of metals from distributed feedback algorithm,
the Cloudy-HZ-LE clusters that might have stayed simply cool are in fact able to become cold.
This is illustrated for two Cloudy-HZ-LE clusters in Fig. \ref{fig:CD_mergers}, where the central
temperature ratio is shown along with major merger events over time.
Note that the mass-scaling relation used to calculate T$_{\mathrm{vir}}$ elsewhere in this paper is not
used in this figure because at early times the mass and redshift of the clusters falls outside the
applicable range of the scaling relation.
Instead, T$_{\mathrm{vir}}$ is calculated by finding the spectroscopic-like temperature inside $R_{200}$,
but with the core removed.
At z=0 this T$_{\mathrm{vir}}$ is roughly similar to the mass-scaled T$_{\mathrm{vir}}$.
In Fig. \ref{fig:CD_mergers} the cluster that ends up warm at z=0 is cool from z$\approx$3 to z$\approx$1.3,
but then undergoes two mergers in quick succession around z$\approx$1.3.
After a period of mixing, over which the core temporarily drops in temperature,
the central temperature rises and never re-cools.
In contrast, the cold cluster experiences no major mergers early in its formation,
which allows the cold core to be established.
It does experience several mergers later in its evolution,
but they happen after the cold core has become established and too gravitationally bound
to be destroyed by the energy of the various mergers.
We stress that while this scenario of the simulated cool/cold-core clusters surviving late major mergers may be inconsistent with observations
that show clusters with recent major mergers lack cool cores \citep[e.g.,][]{2011A&A...532A.123R},
late mergers impacting CCs is not a requirement of our model.

\subsection{Metallicity}\label{sec:MetalComp}

\begin{figure}[htbp] 
   \centering
   \includegraphics[width=0.45\textwidth]{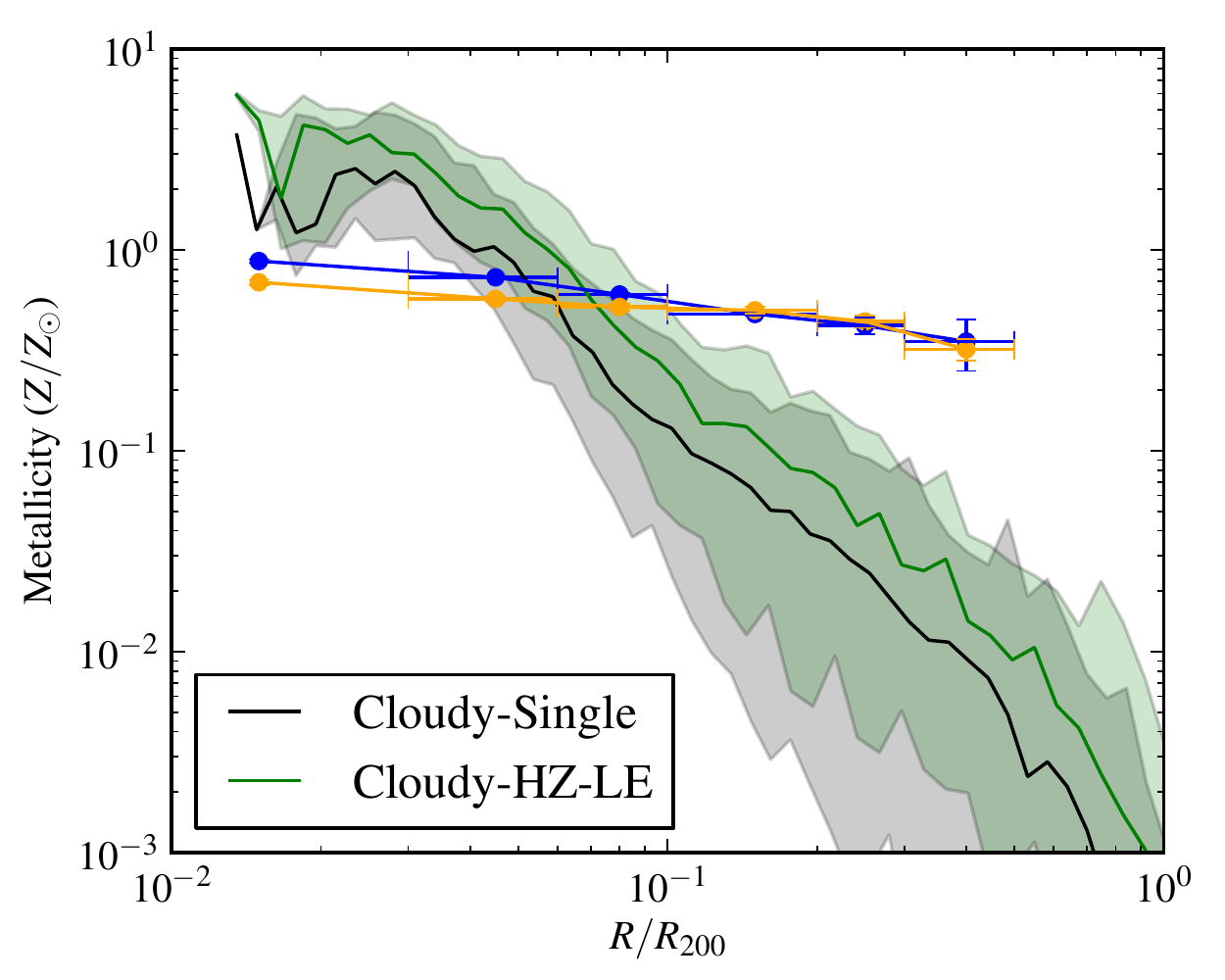}
   \caption{Mean Cloudy-Single and Cloudy-HZ-LE metallicity profiles (solid lines) and $1\sigma$ scatter
   (semi-transparent shaded regions). Observed data from \citet{2011A&A...527A.134M} is shown with error bars;
   blue for clusters with a cD galaxy, and orange for those without.}
   \label{fig:CS-CD_metal_prof}
\end{figure}

We have just shown that simulations using RS (metallicity-independent) cooling fail to reproduce the observed central quantities
due to unphysically high levels of cooling early in the cluster evolution.
Unfortunately, using a metal-dependent cooling algorithm does not result in realistic clusters either.
It is known that a bimodality exists between CC and NCC clusters in which NCC clusters exhibit
a flat metallicity profile towards the center while CC clusters show an enhancement in the core 
\citep{2001ApJ...551..153D, 2004A&A...416L..21B, 2007ApJ...666..835B, 2011MNRAS.413.2467J}.
Clearly, gas metallicity is a very important quantity to examine when attempting to understand CC/NCC clusters,
and in Fig. \ref{fig:CS-CD_metal_prof} we compare the metal profiles
of both \cloudy\ simulations to observations \citep{2011A&A...527A.134M}.
These observations, done with the using XMM-Newton satellite, estimate metallicity by measuring
the strength of iron lines as  function of radius in 28 clusters,
20 of which are also in \citetalias{2010A&A...513A..37H} sample.
Fig. \ref{fig:CS-CD_metal_prof} mirrors the earlier finding that the Cloudy-HZ-LE simulation
has higher mean metallicity than Cloudy-Single,
but it also shows two important differences between simulations and observations.
First, the central ($\lesssim 0.07 R_{200}$) metallicities of both \cloudy\ simulations are far too high,
and second, the slopes of the profiles are far too steep with too little metallicity outside
$\gtrsim 0.07 R_{200}$.
Both effects have been noted before in simulations with star formation + cooling and
without AGN feedback \citep[e.g.,][]{2006MNRAS.366..397S}.

Both effects are caused by star formation that is too centrally concentrated.
Nearly all of the star formation occurs inside $0.1 R_{200}$ in the region of the
cluster with the highest gas density,
which implies that essentially all of the thermal feedback is contained within this region.
This effect is still present when distributed feedback is used,
since the region over which feedback takes place for each star particle is a cube $\approx47 h^{-1}$ kpc on a side -- 
significantly smaller than the volume contained within $0.1 R_{200}$.
Given the high density of the gas, cooling takes place rapidly,
lowering the entropy of metal-enriched gas and keeping it contained within the center of the cluster.
In principle, star formation in galaxies near the outskirts of clusters could result in substantial additional enrichment of the gas.
In practice, however, the relatively low resolution of these calculations prohibits star
formation in all but the largest galaxies in a cluster, resulting in highly centralized star formation. 
As a result of these two factors, the metallicity profile in all of our simulated galaxy clusters is heavily tilted,
with the inner regions of the clusters having far too much metal, and the outer regions having too little.

\subsection{The Effect of Cold Gas on Observables}\label{sec:coldgas}

The orbiting X-ray observatories Chandra and XMM-Newton have little to no sensitivity below $\approx 0.5$ keV,
which means that they are unable to accurately detect gas colder
than a few million K \citep{2003SPIE.4851...28G, 2001A&A...365L...1J}.
The simulated cold-core clusters discussed earlier in this section contain gas at their center well
below this temperature threshold, which means that the cold gas would be invisible to the observatories.
Therefore, to more accurately model observables as they might be measured by the aforementioned observatories,
it is appropriate for the calculations of the
observables to remove the cold central gas from consideration \citep{2011ApJ...731L..10N}.
As described in Section \ref{sec:TspecDef}, mock X-ray measurements like $T_{spec}$
are unphysical at these low temperatures. Calculations of the
metallicity and other X-ray derived quantities from simulations are
likewise skewed if this non-X-ray emitting gas is included.

The removal of cold gas from the calculation is accomplished for
each cluster by first finding the total central X-ray emissivity in the 0.5-7.0 keV bandpass.
Next, we find the temperature for which gas below that temperature accounts for only 5\% of the total emissivity.
We then eliminate the gas below this threshold from our calculations of all three main observables.
The mean temperature cut across all clusters is roughly $1-2\times10^7$ K,
which is very similar to the lower temperature limit of the observatories quoted earlier.
We show the result of these temperature cuts for two simulations, Cloudy-Single and Cloudy-HZ-LE,
in Fig. \ref{fig:cc-cuts2}(e, f).
We note that when we perform this same temperature cut procedure on the other simulations
(not plotted),
the differences in the observables are negligible because those simulations
contain very little or no gas below $1-2\times10^7$ K in the centers of the clusters.

With the temperature cuts, the cold-core clusters disappear, and the simulations appear more similar to observations
(see panels (e, f) of Fig. \ref{fig:cc-cuts2}).
In fact, Cloudy-HZ-LE without cold gas produces a distribution of central cooling times
just on the cusp of being likely to be consistent with observations ($p$=0.01).
However, we argue that this improvement is of mixed value for analyzing these simulations.
While the cold-core clusters do appear more realistic using the temperature cuts,
it hides the true nature of the cold gas which is unlikely to be physical.
For example, \citet{2010ApJ...719.1844H} find using UV GALEX observations of CC clusters
(some of which are in the  \citetalias{2010A&A...513A..37H} sample)
that star formation rates in the inner 50-100 kpc are generally well under 1 M$_{\odot}$/yr.
The low rate of star formation in observed CC clusters indicates that there is not a
large amount of very cold gas present in the centers of CC clusters.
Because the cold gas is therefore unlikely to be physical, and clusters without the unphysical cold gas are unaffected by the temperature cuts,
the {\it inclusion} of the cold gas in our calculations is actually more informative for purposes of understanding the simulations
and identifying problematic clusters.
Therefore, we will include the cold gas in our calculations for the remainder of our analyses.

\section{The Effect of Varied Feedback and Cluster Mass on Observables}\label{sec:AdjustFeedback}

The results presented in the previous demonstrate that simply changing radiative
cooling methods and the way that stellar feedback is deposited into the intracluster medium cannot
produce physically-reasonable simulated clusters that agree with observations.
Therefore, we move on to varying the quantities of feedback into the ICM from supernovae,
which are remaining free parameters in our model.
The results from Section \ref{sec:FirstFourSims} indicate that the RS cooling results
(using a metallicity-independent cooling table) fail to produce realistic clusters,
which leaves the \cloudy\ (e.g., metallicity-dependent cooling) simulations as options for a platform for further exploration.  
It is arguable which of the two \cloudy\ cooling simulations better reproduce observations,
but we choose to use Cloudy-HZ-LE as the basis for further testing because
it employs both metallicity-dependent cooling and the distributed stellar feedback algorithm.
The \cloudy\ cooling has no adjustable parameters, leaving only the feedback parameters
$\epsilon_E$ and $\epsilon_Z$.

Because the central metallicity of Cloudy-HZ-LE is so much higher than observations,
and it is clear that the high metallicity has a large effect on the observable properties of the simulated cluster cores,
we run three simulations in which we lower the metal feedback parameter ($\epsilon_Z$) by a factor of five,
and we vary the value of the energy feedback parameter ($\epsilon_E$) (see Table \ref{table:SimParams}).
These three simulations are labeled ``Cloudy-LZ-LE'' (``Cloudy-Low Metal Feedback-Low Energy Feedback''),
``Cloudy-LZ-ME'' (Medium Energy Feedback), and
``Cloudy-LZ-HE'' (High Energy Feedback).
The first of the three has the same value of $\epsilon_E$ as Cloudy-HZ-LE, while the next two
increase $\epsilon_E$ by a factor of 8 and 20, respectively.

\begin{figure}[htbp] 
   \centering
   \includegraphics[width=0.45\textwidth]{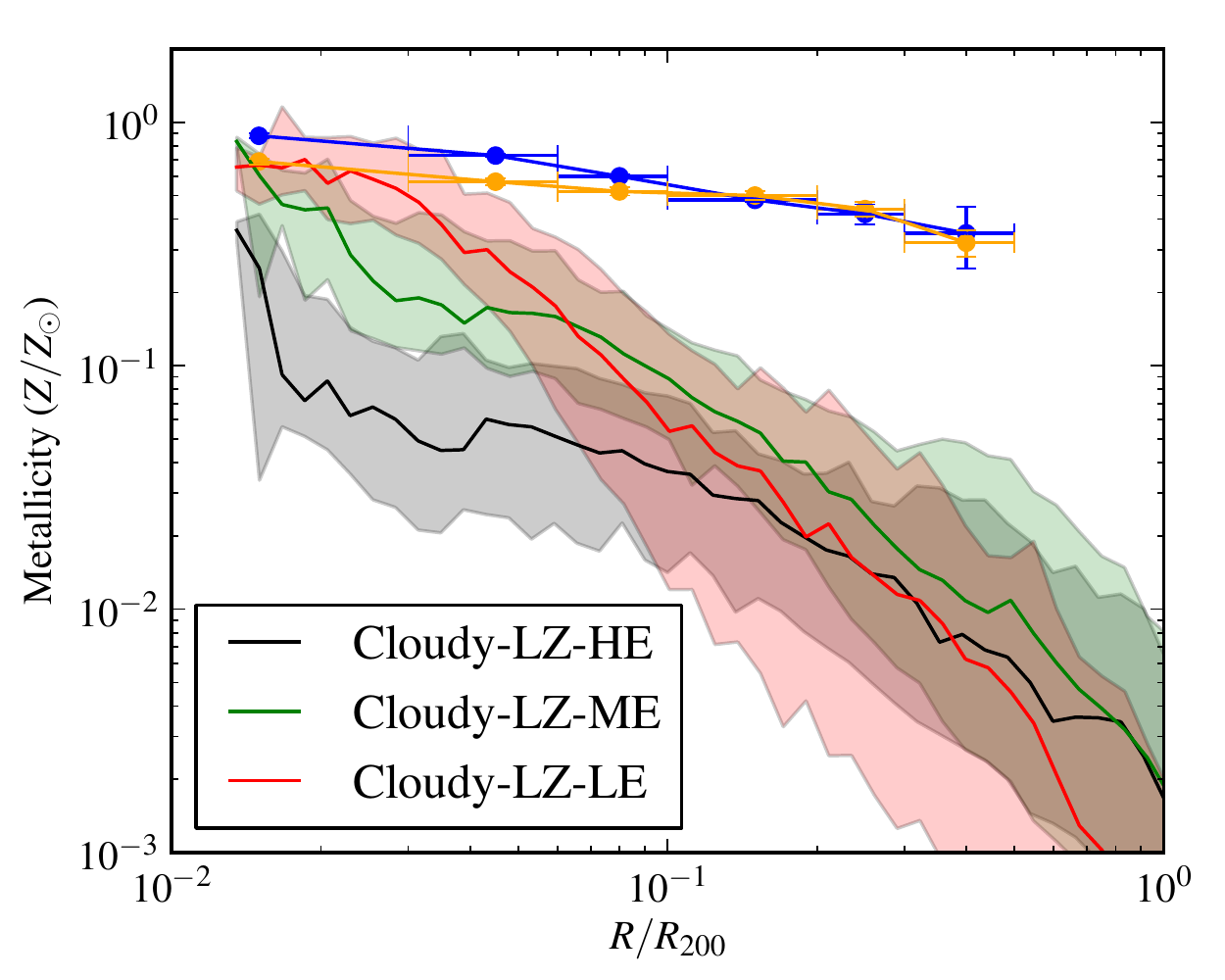}
   \caption{Mean metallicity profiles for all three low metal \cloudy\ runs (solid lines)
   and $1\sigma$ scatter (semi-transparent filled regions).
   Observed data from \citet{2011A&A...527A.134M} is shown with error bars;
   blue for clusters with a cD galaxy, and orange for those without.}
   \label{fig:3LowMet_metal_prof}
\end{figure}

Figure \ref{fig:3LowMet_metal_prof} shows that the lower $\epsilon_Z$ has
dramatically reduced the central metallicities in all three simulations.
Indeed, inside roughly $0.05R_{200}$ (which includes our defined central region),
the Cloudy-LZ-LE metallicities agree fairly well with observations.
However, outside that region the metallicity still declines far too rapidly for all three simulations.
This is not surprising since, to first order, we have simply reduced the amount of metal
that is produced overall in the simulation, but not the locations or timing of star formation.

\begin{figure*}[htbp] 
   \centering
   \includegraphics[width=1.0\textwidth]{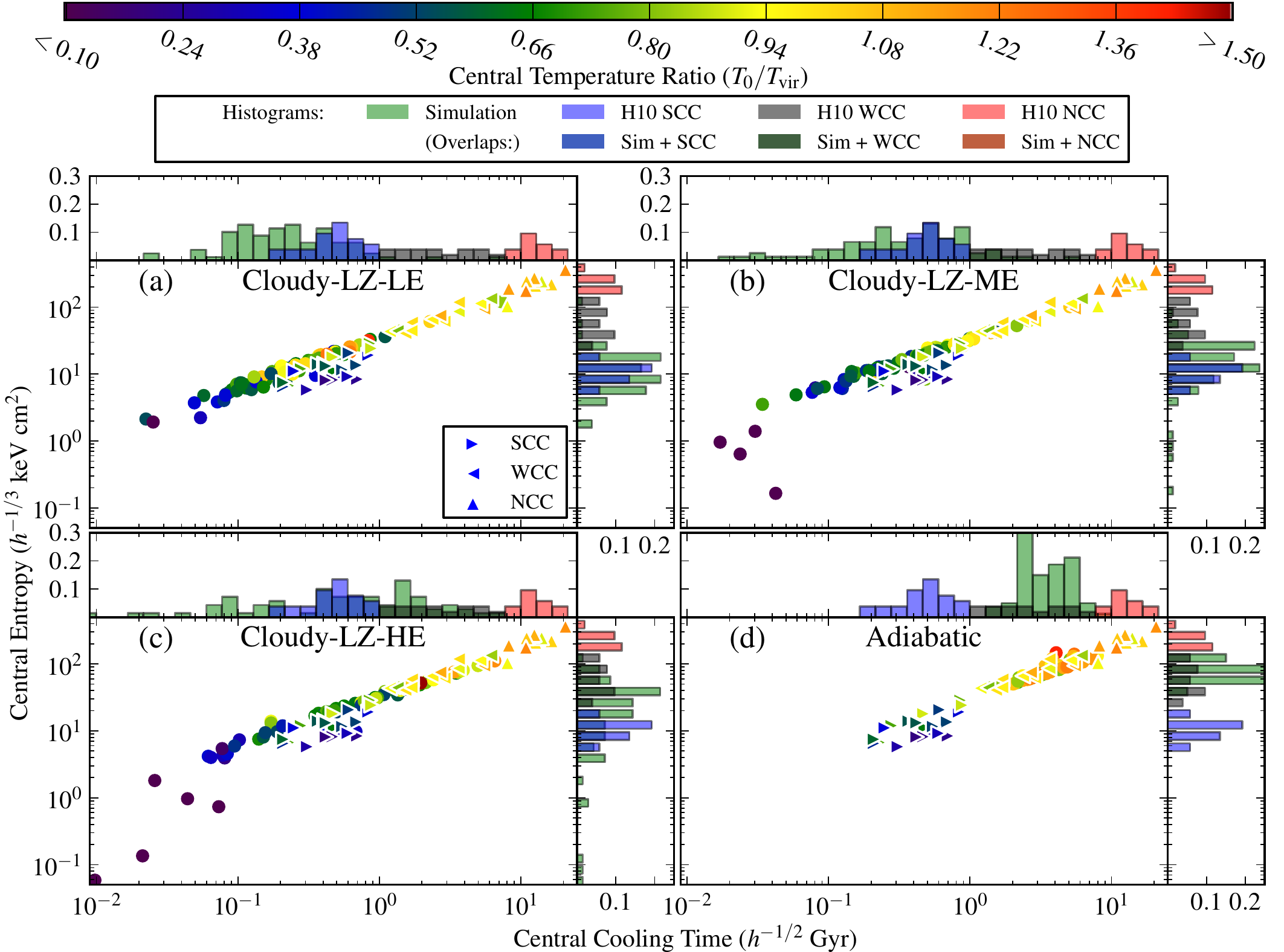}
   \caption{Plots and histograms of central cooling time and entropy for simulated and observed clusters;
   similar to Fig. \ref{fig:cc-cuts2}. 
   See \S\ref{sec:FirstFourSims} for a detailed description of the aspects of this figure.
   Gas temperature cuts have not been applied to the calculations in this figure.
   The ``Adiabatic'' simulation is discussed in Section \ref{sec:Energetics}.}
   \label{fig:cc-cloudy}
\end{figure*}

In Figure \ref{fig:cc-cloudy} we show the results of varying the feedback energy on our observables.
The figure shows that the lower and more realistic central metallicities have a profound effect
on the observables.
From Cloudy-HZ-LE (Fig. \ref{fig:cc-cuts2}(d)) to Cloudy-LZ-LE, the very low entropy tail of Cloudy-HZ-LE has vanished,
and in its place the central entropy distribution is now peaked solidly, but too strongly,
in the observed SCC range.
Additionally, lowering the metal feedback has significantly altered the distribution of central temperature ratios
(Fig. \ref{fig:hist-Tdrop}(e--g)), and brought them closer to agreement with observations.
In particular, the KS test between observations and the Cloudy-LZ-LE central temperature ratio distribution gives $p=0.03$,
which is the best match out of all the distributions presented in this paper.

The other two low-metallicity runs show that by adjusting the feedback parameters, we are able
to produce more realistic clusters, at least as measured by these three observables.
In particular, except for a few simulated cold-core clusters (which we discuss below), the agreement
between the observed SCC+WCC distributions of central cooling time and of entropy and
the Cloudy-LZ-HE clusters is visually better than any of the other simulations.
As in all other simulations, KS tests produce $p$-values less than 0.01, but this is largely due to our inability to produce
clusters with NCC-like cooling times and entropies.

\begin{figure}[htbp] 
   \centering
   \includegraphics[width=0.45\textwidth]{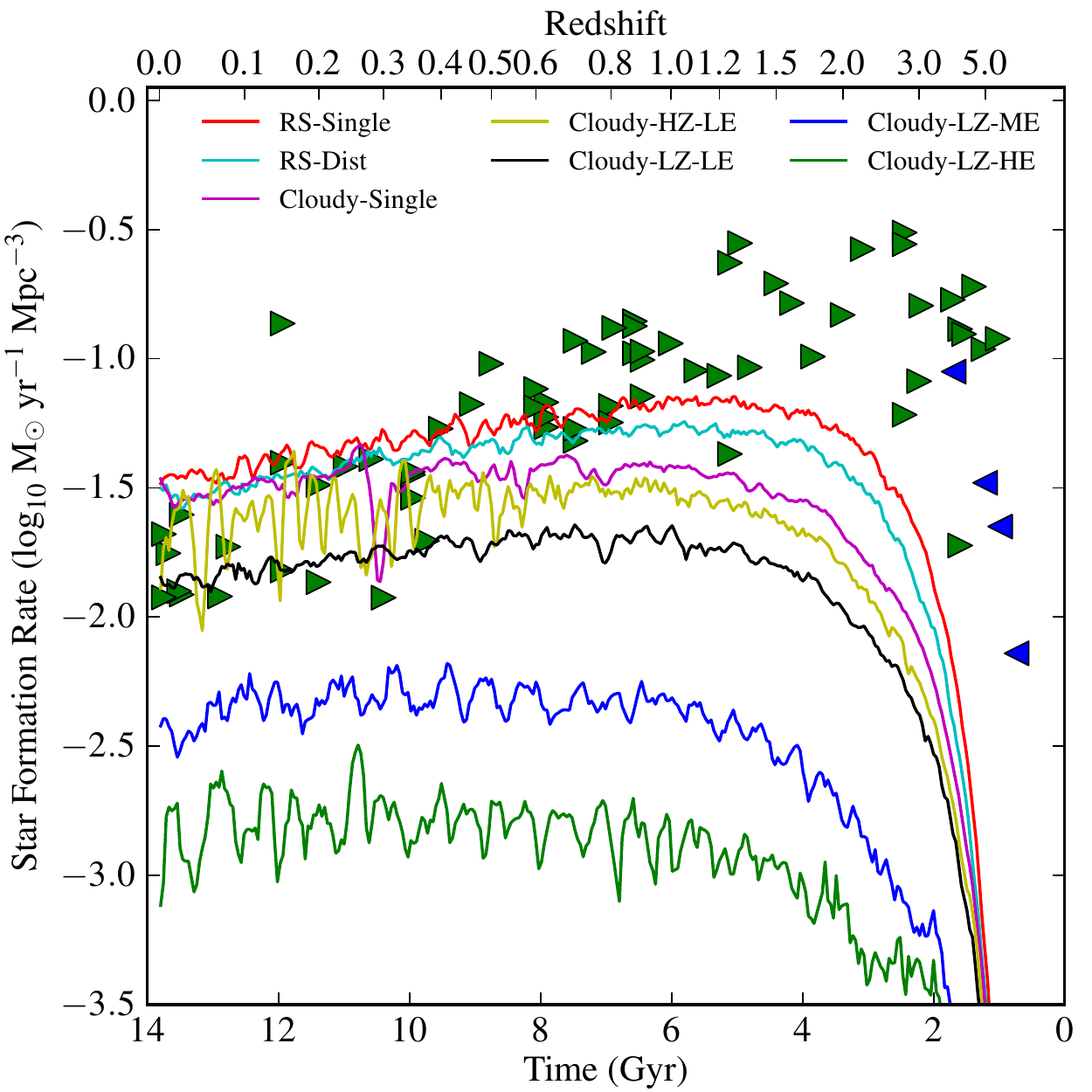}
   \caption{Star formation rates over time for each simulation.
   We plot observed rates as a visual guide, but we do not expect concordance.
   The simulated rates have been smoothed by averaging over a moving window $\sim200$ Myr wide.
   The blue left-pointing triangles are observed rates from \citet{2007ApJ...670..928B},
   and the green right-pointing are from \citet{2004ApJ...615..209H, 2007ApJ...654.1175H}.
   Error bars have been omitted from the observations for clarity.}
   \label{fig:SFR}
\end{figure}

These three low metallicity simulations produce a few interesting trends that allow us to highlight
the interconnectedness of star formation, gas enrichment, and gas cooling that is at the core of our
inability to produce a realistic distribution of clusters.
First, the simple act of reducing the metal feedback from Cloudy-HZ-LE to Cloudy-LZ-LE noticeably shifts
the distributions of cooling times and entropies to lower values.
This is a direct result of the fact that lowering the amount of metals in the gas reduces gas cooling,
and this in turn reduces the star formation rate (see Fig. \ref{fig:SFR}).
Lower rates of star formation reduces the amount of energy feedback available for raising
the central cooling times and entropies.
We discuss Figure \ref{fig:SFR} in more detail
in Sections \ref{sec:LackRes} and \ref{sec:Interconnect}.

Second, as $\epsilon_E$ is increased, the bulk of the central cooling times and entropies moves to greater values,
and becomes more in line with the SCC+WCC portion of the observed distributions.
This is not a surprising effect.
Raising the energy feedback of star formation acts to heat up the gas, which raises cooling time,
and also increases the entropy.
Note that despite increasing $\epsilon_E$ by a factor of 20, we still fail to produce fully-NCC
(i.e. in the NCC range of all three central quantities) clusters.

\begin{figure}[htbp] 
   \centering
   \includegraphics[width=0.45\textwidth]{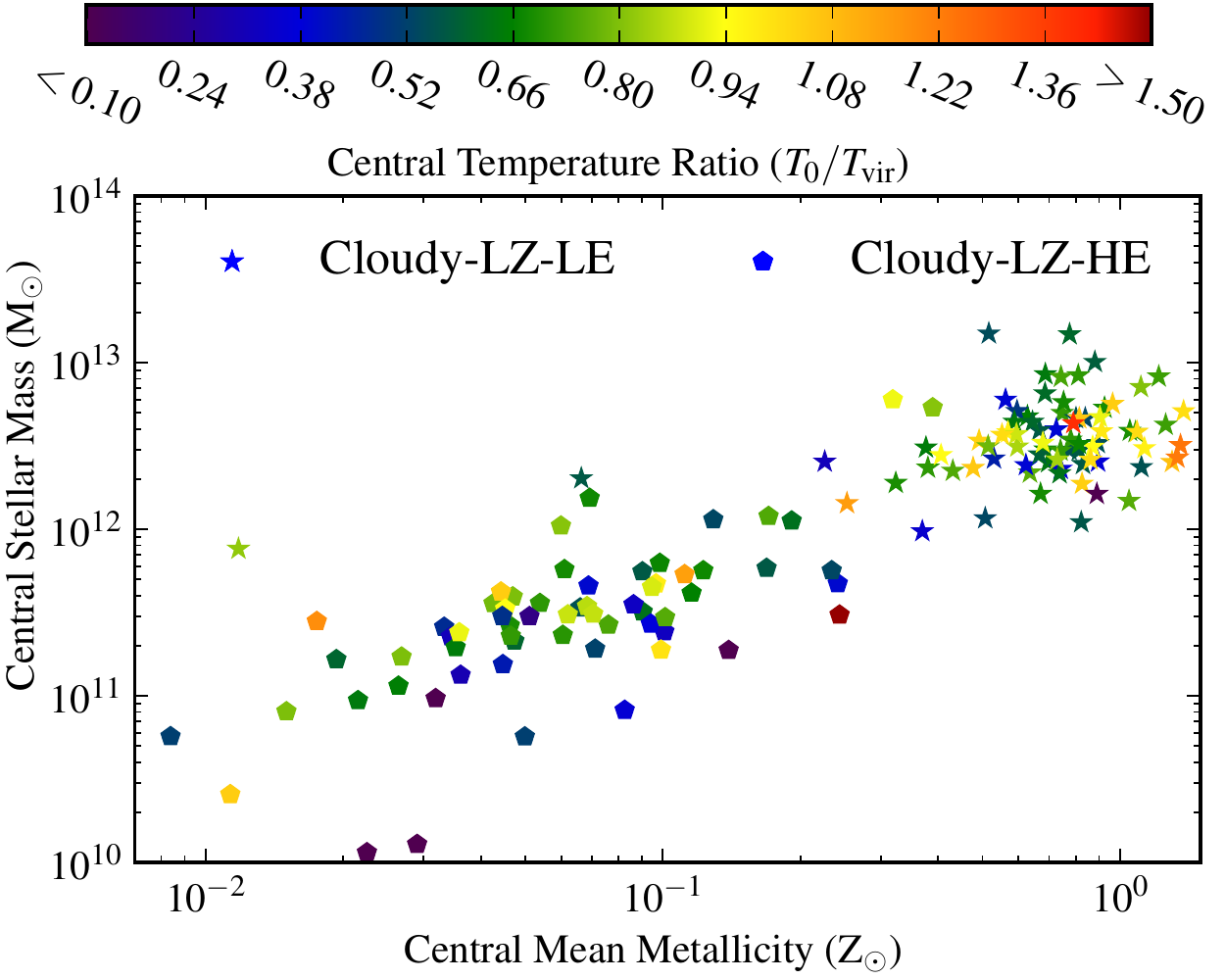}
   \caption{The central mean metallicity versus total central stellar mass
   for Cloudy-LZ-LE and Cloudy-LZ-HE.}
   \label{fig:SM_Z}
\end{figure}

Curiously, {\it increasing} the energy feedback also increases the number of
very low entropy, cold-core clusters from Cloudy-LZ-LE to Cloudy-LZ-HE.
In contrast to the Cloudy-HZ-LE cold cores, which are primarily due to the high metallicity of the ICM
in the center of the cluster,
the increase in the number of cold cores seen in the Cloudy-LZ-HE simulation are due to a combination of
low metallicities and low amounts of formation of stars and their associated feedback energy.
Figure \ref{fig:SM_Z} shows that the coldest Cloudy-LZ-HE clusters
also have relatively low total stellar masses.
In the higher energy feedback run, 
it appears that the high thermal feedback of a given generation of stars inhibits the formation
of the next generation of stars, forming a feedback loop.
In some clusters this feedback loop is very strong, and despite (and due to) the high value of $\epsilon_E$,
the star formation is not enough to warm up the core.
This suppression of star formation as feedback energy is increased is also illustrated in Fig. \ref{fig:SFR}.

\begin{figure}[htbp] 
   \centering
   \includegraphics[width=0.45\textwidth]{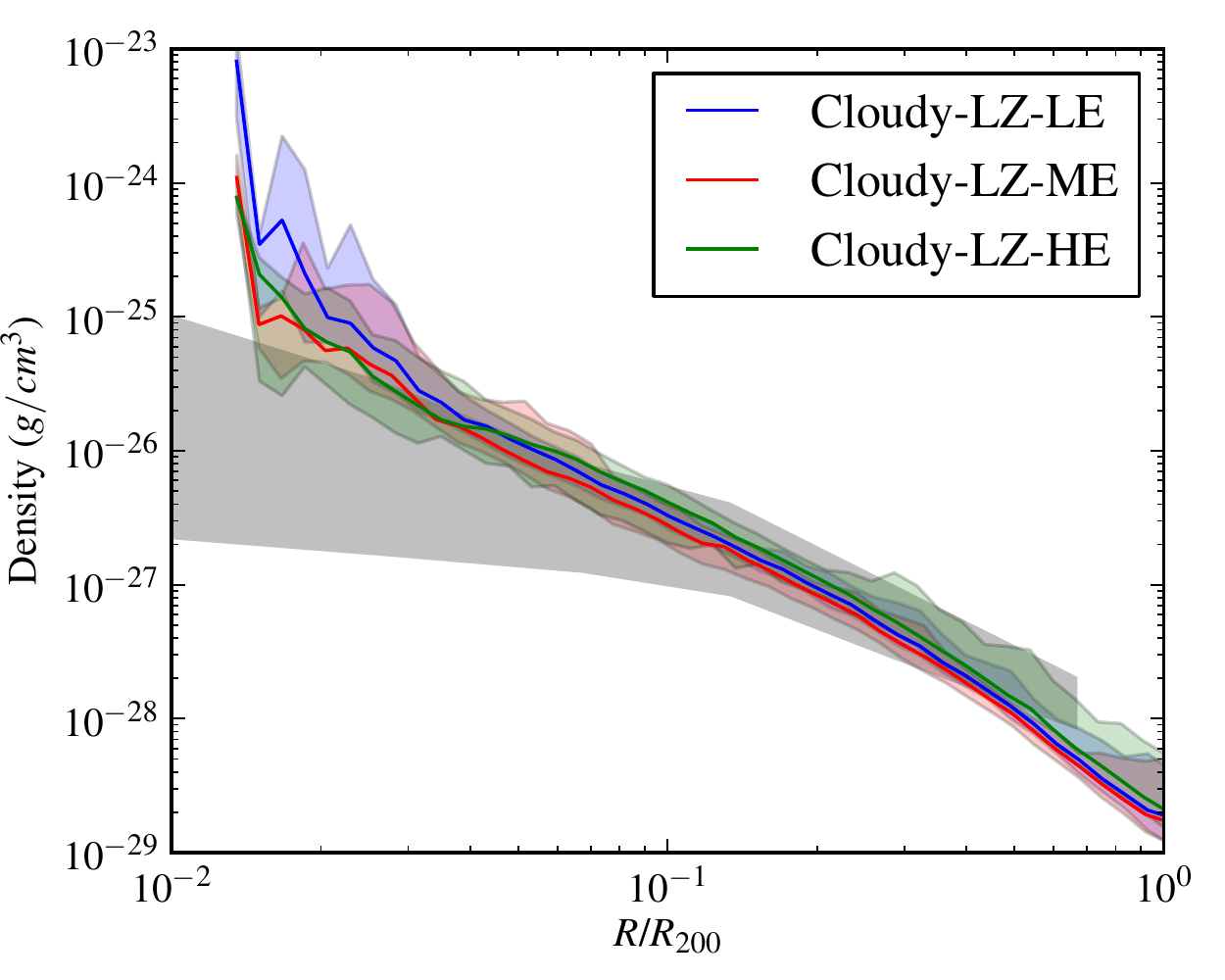}
   \caption{Mean density profiles for each simulation as labeled (solid lines) with $1\sigma$ scatter
   (semi-transparent filled regions). The range of observed densities from \citet{2011A&A...536A...9P}
   is over-plotted in grey. These density profiles are essentially unchanged from the 
   earlier simulations using \cloudy\ cooling (Fig. \ref{fig:All4DensityProfiles}).}
   \label{fig:All4LZDensityProfiles}
\end{figure}

\begin{figure}[htbp] 
   \centering
   \includegraphics[width=0.45\textwidth]{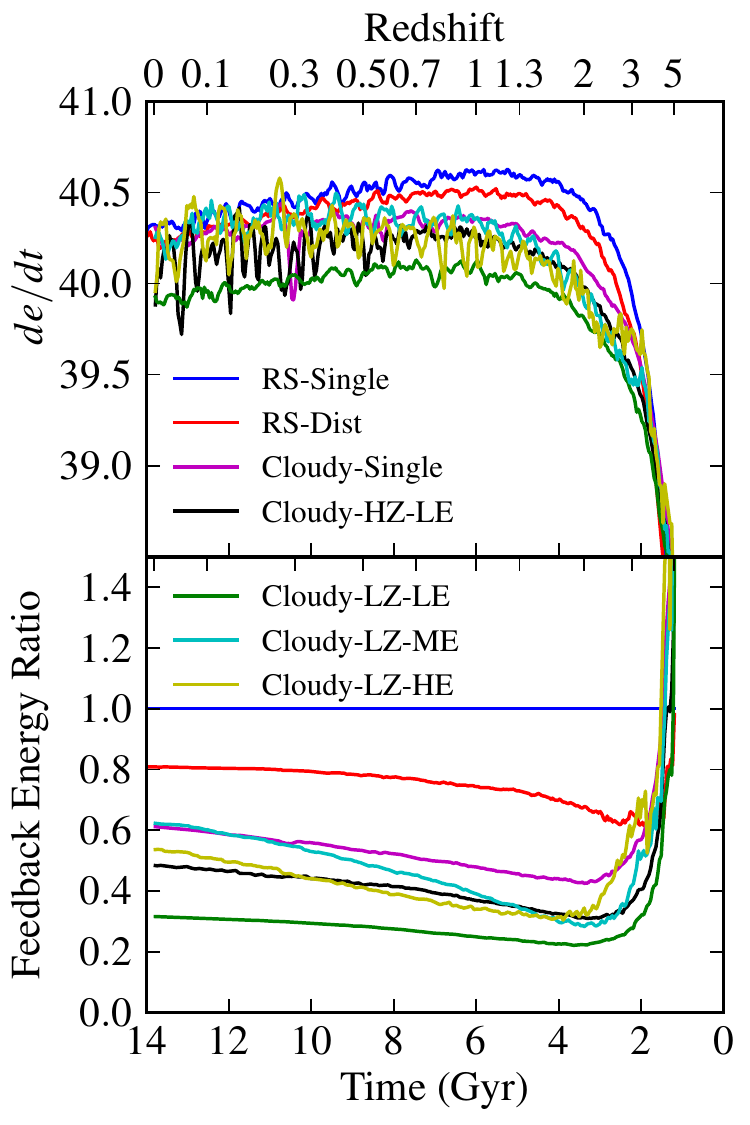}
   \caption{
   {\it Top panel:} Instantaneous feedback energy over time
   averaged over a
   moving window $\sim200$ Myr wide. The units are in $\log_{10}($ergs s$^{-1}$ Mpc$^{-3})$.
   {\it Bottom panel:} Full-box total energy feedback integrated over time, as a function of time,
   for each simulation divided by the same for RS-Single (hence RS-Single in blue is a flat line).
   }
   \label{fig:SFR-Energy}
\end{figure}

Third, while the central cooling times and entropies shift as the thermal feedback is modified,
the central temperature ratios (Fig. \ref{fig:hist-Tdrop}(e--g)) do not change nearly as much
(except for the addition of the few cold-core clusters at the cold end of the distributions).
The reasons why the central temperatures do not change very much with $\epsilon_E$,
while the cooling times and entropies do, are complex.
Figures \ref{fig:3LowMet_metal_prof} and \ref{fig:SFR} show that the increase in $\epsilon_E$
lowers both the central metallicities and star formation rates.
This results in lower cooling rates which prevents central densities from becoming quite as high
(see \ref{fig:All4LZDensityProfiles}).
Keeping roughly equal temperatures, but having lower densities,
results in longer cooling times and higher entropies.
Across the three simulations the temperatures stay roughly equal because, despite the factor of 20 increase in
the amount of thermal energy returned to the intracluster medium per solar mass of star formed,
the amount of feedback energy returned to the gas stays strikingly
constant (Fig \ref{fig:SFR-Energy}).
From Cloudy-LZ-LE to Cloudy-LZ-HE, there is less than a factor of two increase in total
energy deposited in the gas.
Figure \ref{fig:SFR-Energy} is discussed in more detail in
\S\ref{sec:Energetics}.

\subsection{Higher Mass Clusters}\label{sec:HighMassClusters}

Above and in \S\ref{sec:FirstFourSims} we removed 12 clusters from the \citetalias{2010A&A...513A..37H} sample
because their masses are higher than the most massive ones in our simulations.
Due to their deeper potential wells and likely richer history of mergers,
it is reasonable to expect that higher-mass clusters would tend toward having NCCs when compared to less massive clusters.
In order to test if higher-mass clusters behave differently than the ones already discussed,
we ran a suite of follow-up simulations of higher-mass clusters.
The higher mass clusters do not show any meaningful changes in their physical observables when compared
to their lower mass analogues, indicating that simply producing higher mass clusters does not
produce more realistic clusters.

\section{Discussion}\label{sec:Discussion}

In this paper, we have shown that \cloudy\ (metallicity-dependent) cooling and distributed thermal feedback can be
used to create simulated galaxy clusters with somewhat more realistic ICM properties than in previous generations of simulations.
Even our best models, however, fail to accurately reproduce the observed cluster properties --
in particular, the distribution of clusters into cool core and non-cool core clusters and related properties:
the central entropy and cooling time of the intracluster gas,
the ratio in temperature of this central gas from the cluster's virial temperature,
and the distribution of metals within the cluster itself.
In particular, virtually none of our simulated clusters fall into the non-cool core ranges of central cooling time and entropy.
In this section, we discuss possible reasons for the continued difference between our simulation clusters and real galaxy clusters.

As discussed in the introduction to this paper,
there is a large body of numerical work that shows that AGN can help to alleviate
some of the problems exhibited by our simulations,
including regulating star formation and promoting the distribution of metals.
In Section \ref{sec:Con} we discuss a sub-grid feedback method that we believe will allow us
to address the some of problems discussed in this section and should allow us to
include AGN feedback to our clusters as well.

\subsection{Lack of Resolution}\label{sec:LackRes}

Due to finite computational resources and the need to simulate a large volume of the Universe,
our galaxy cluster simulations have relatively poor
mass and spatial resolution compared to the current cutting-edge in galaxy or single-cluster
(e.g., \citealt{2012ApJ...747...26L}) formation simulations.
This has a negative effect on the accuracy of our model, and is
especially true at early times, when cooling and star formation is taking place in real galaxies with masses
below the resolution of these simulations.
For example, at z$\approx6$ the dominant galaxy population in terms of feedback is $\approx10^{10}-10^{11}$ M$_\odot$
\citep{2007ApJ...670..928B, 2011ApJ...735L..34G},
and at z=2 it is $\approx10^{12}$ M$_\odot$ \citep{2009ApJ...696..620C}.
Given that the dark matter particle mass in our simulations is $7.8\times10^9$ M$_\odot$,
we simply cannot resolve galaxies at high z, and therefore we do not make stars at that epoch.
This is clearly illustrated in Fig. \ref{fig:SFR}.
Furthermore, in \Enzo, as in hydrodynamical simulations in general, gas cooling within halos only becomes significant once 
once the gravitational potential is well-resolved, and thus can compress gas to high densities and form stars.
From a practical standpoint, this means that a halo must contain a relatively large number of dark matter
particles before star formation can occur -- substantially more than is required to simply register the existence of a halo.
This is illustrated by the fact that in these simulations
the average halo has a mass of $7.3\times10^{12}$ M$_\odot$ when a star is first formed inside it,
which corresponds to $\approx1,000$ dark matter particles.
Because the halos that form stars for the first time are so massive,
star formation occurs later in time and in objects with deeper potential wells (and thus denser gas) than is typical in the real Universe.
The denser gas is able to cool more readily, which means that the thermal feedback from star formation
does not warm up gas as effectively, and the overall result is an excess of cooling and a deficit of entropy at low redshifts.
The injection of metals into the intergalactic medium is similarly affected by our lack of resolution,
with metal-enriched gas being more strongly concentrated than is observed.
This leads to an unfortunate feedback loop -- stars form too late and in over-dense regions,
which are then over-polluted with metals.
This excess of metal in the gas causes cooling to occur too rapidly,
reducing or eliminating the pressure gradients that would drive metal-enriched gas to large cluster radii.

The star formation rate in our simulations is illustrated in Figure \ref{fig:SFR}.
Observed data from \citet{2007ApJ...670..928B} (from the HST HUDF and GOODS fields)
and \citet{2004ApJ...615..209H, 2007ApJ...654.1175H} (combined from the literature)
are included in the figure for comparison.
This illustrates that all of the calculations discussed in this paper differ substantially from observed rates at z$>$0.3,
and in some cases do so quite dramatically.
Of course, as mentioned above, the simulations do not resolve the galaxies
that are responsible for a large fraction of star formation,
and therefore it is unreasonable to expect precisely matching rates.
Rather, the figure confirms that star formation begins later than the cosmic mean,
which is exactly the opposite of what one would expect from observations of galaxy clusters,
where the bulk of stars form \textit{earlier} than the cosmic average due to clusters being substantially
over-dense regions at high redshift.

\subsection{Interconnectedness of Star Formation and Feedback}\label{sec:Interconnect}

A striking feature of the different stellar histories in Fig. \ref{fig:SFR} is how
much they change as the models for radiative cooling and subgrid star formation are adjusted.
Every deliberate choice we made to improve the distributions of
our three core observables has, in fact, caused the timing of the peak rate of star formation to
occur later and later.
Recall that this is contrary to what is desired, which is a cluster star formation rate peak that occurs
earlier than the overall universal peak.
This effect, and the effects noted earlier of adjusting $\epsilon_E$ and $\epsilon_Z$,
illustrate the interconnectedness of star formation and feedback in our simulations.

An example of this effect are the low-metallicity runs discussed in \S\ref{sec:AdjustFeedback}.
The first run, Cloudy-LZ-LE, that only reduces the metal feedback, results in clusters that have systematically
low central entropies and cooling times.
Increasing the energy feedback does somewhat ameliorate those particular discrepancies,
but the result is an even less realistic history of star formation.
It is clear that the {\it star formation $\rightarrow$ metal feedback $\rightarrow$ gas cooling} cycle is a self-reinforcing cycle.
A logical change to one aspect of our subgrid stellar model results in other aspects
drifting further from the observational results we are attempting to match.
As mentioned previously, the cold-core clusters seen in Cloudy-LZ-HE are evidence of an incorrectly modeled
cycle of star formation and metal enrichment.
This same behavior exists in all the other simulations discussed in this work,
and it is impossible to disentangle a change in one part of that cycle without
affecting all the others, often negatively.

More broadly, the fundamental challenge with our subgrid models for star formation
and feedback is their strong dependence on the mass and spatial resolution of the simulation,
as was discussed in Section \ref{sec:LackRes}.
Using our current technique, we will not be able to correctly resolve the early evolution of the
galaxy populations in proto-cluster regions,
which limits our ability to bring the observable properties of our simulated clusters in line with real galaxy clusters.
Given this unresolvable issue,
it is clear that a fundamentally different approach needs to be taken,
which we will elaborate upon below.

\subsection{Energetics}\label{sec:Energetics}

Our lack of fully-NCC clusters indicate that
we are not correctly applying thermal feedback to these simulations.
As an experiment, we run a simulation
that has identical initial conditions and simulation parameters to the other calculations described in this paper,
except that radiative cooling, star formation, and stellar feedback are all disabled.
We label this calculation ``Adiabatic,'' and the results are shown in
Figures  \ref{fig:hist-Tdrop}(h) and \ref{fig:cc-cloudy}(d).
The only source of cooling in this calculation is from adiabatic expansion of the plasma,
and the only sources of heating comes from adiabatic compression and from shock heating,
powered by gravitational potential energy.
As can be seen in Figures \ref{fig:hist-Tdrop}(h) and \ref{fig:cc-cloudy}(d),
the Adiabatic simulation results in warm galaxy cluster cores, with the majority of
simulated clusters solidly in the NCC range
of the central temperature ratio distribution (Fig. \ref{fig:hist-Tdrop}(h)).
However, the central cooling times (as before, calculated using a cooling curve based on an optically thin plasma with a metallicity of 0.5)
and entropies are only in the WCC ranges (Fig. \ref{fig:cc-cloudy}(d)),
which shows that simply heating the core to NCC-like temperatures is not enough to produce
fully-NCC clusters.
The Adiabatic results also highlight the requirement that clusters include both heating and cooling
because without the early cool phase of clusters, some of which become may NCC clusters following mergers,
it is impossible to create the full spectrum of cluster types.

We now turn to Figure \ref{fig:SFR-Energy}, which demonstrates the insensitivity of the observables
in our simulations to our choice of cooling algorithm and magnitude of both metal and thermal feedback. 
The instantaneous energies are calculated for each simulation by convolving the star formation rate
(Fig. \ref{fig:SFR}) with Eqn. \ref{eqn:E} and the value of $\epsilon_E$ (Table \ref{table:SimParams}).
Following the rates in the top panel,
the maximum instantaneous ratio between any two simulations is
under an order of magnitude.
The bottom panel is perhaps even more interesting.
This shows that across all seven simulations the total feedback energies deposited
in the gas are remarkably similar.
From the highest total in RS-Single, to the lowest in Cloudy-LZ-LE,
there is only about a factor of three change in the amount of energy deposited,
which stands in stark contrast to the large changes in physics modules
and feedback parameters.
This shows that using our current tools it is very difficult to change the energy feedback history,
and that it is likely very challenging to simulate the required energy feedback history to produce realistic clusters
with the parameters we are able to adjust.

Despite the assertions above about little energetic differences between simulations,
there {\em are significant differences} in the observables across the simulations.
The fact that even small changes in energetics lead to major shifts in
(especially) entropy and cooling time highlights
the importance of accurately modeling energy physics in cluster simulations.
As mentioned before, we discuss in \S\ref{sec:Con} a method that should allow us to more accurately model
the energetics in clusters and reproduce more realistic results.

\section{Conclusions and Future Work}\label{sec:Con}

Where are we along the road to making realistic galaxy clusters?
In Table \ref{table:WinLose} we summarize qualitatively
the successes and failures of our seven main simulations, basing our statements
on the data contained in Figures  \ref{fig:cc-cuts2}, \ref{fig:hist-Tdrop}, and \ref{fig:cc-cloudy}.
Since no simulations produced fully-non-cool clusters (as defined by the measured cooling times and entropies),
we omit that as a comparison from the table,
and note that it is a failure of all the simulations.
However, we also note that all simulations, except those using metal-independent RS cooling, did produce a substantial fraction of NCC clusters
as measured by central temperature ratio.
Where we feel the judgment is warranted, we have labelled
assessments with either a ``best'' or a ``worst.''
Our assessments of the three main observables illustrate that overall none of the
simulations did particularly well at reproducing the observed distributions.
The simulations using cooling tables that assume a constant, high metallicity
(the RS series of runs) are the furthest from matching observations, primarily due to over-cooling of gas.
Among runs using metal-dependent cooling (the \cloudy\ series of runs),
the simulation with the low metal feedback and high thermal energy feedback (Cloudy-LZ-HE)
overall produces the best match to observations, with reasonable agreement for both cooling time and
decent agreement for both cooling times and entropies,
and central temperatures almost entirely within the observed range,
although the full spectrum of cool-core and non-cool core cluster is not reproduced.

\renewcommand{\tabcolsep}{1pt}
\begin{deluxetable*}{rccccc}
\tablecolumns{6}
\tabletypesize{\scriptsize}
\tablecaption{Simulation Successes and Failures\label{table:WinLose}}
\tablewidth{0pt}
\tablehead{
\colhead{Simulation Label} & \colhead{Cooling Time} & \colhead{Entropy} &
\colhead{Temperature Ratio} & \colhead{Metallicity Profile} &
\colhead{Star Formation}
}
\startdata
RS-Single & Worst, Entirely & Mainly Too Low  & Mostly CCs, & N/A & Best Overall, Too Low \\
& Too Short &  & Too Few NCCs &  & Early, Too Slow Rolloff \\
RS-Dist & Mainly Too Short & Mainly Too Low & Mostly CCs, & N/A & Too Low Early,  \\
&  &  & Too Few NCCs & & Too Slow Rolloff\\
Cloudy-Single & Best SCC+WCC  & Too Peaked Between & Too Many NCCs, & High in Center, & Too Low Early, \\
& agreement & SCCs and WCCs  & Too Few CCs &Profile Too Steep & Too Slow Rolloff \\
Cloudy-HZ-LE & Decent SCC+WCC, & Many Cold-Cores With & Too Bimodal & Worst, Far Too High in & Far Too Low Early, \\
& But A Few Too Low & Too Low Values &  & Center, Profile Too Steep & Too Slow Rolloff  \\
\vspace{1pt} \\
\hline
\vspace{1pt} \\
Cloudy-LZ-LE & Generally Too Short & Mainly SCCs, & Peaked Between the & Best in Center, Low in & Far Too Low Early,  \\
&  & A Few WCCs & CC \& NCC Groups & Periphery, Profile Too Steep & Too Slow Rolloff   \\
Cloudy-LZ-ME & Mainly SCC, & Almost Entirely & Peaked Between the & Low in Periphery, & Far Too Low Early, \\
& A Few WCCs &  WCCs & CC \& NCC Groups &Profile Too Steep & No Rolloff  \\
Cloudy-LZ-HE & Decent SCC+WCC, & Decent SCC+WCC, & Peaked Between the & Low Overall, & Worst, Abysmal Always,  \\
& A Few Too Short & A Few Too Low & CC and NCC Groups & Profile Too Steep & No Rolloff  
\enddata
\tablecomments{The entries are qualitative assessments of how
well the simulations agree with observations.
\vspace{5pt}
}
\end{deluxetable*}

We have also included assessments of the metallicities and star formation histories
of the simulations in Table \ref{table:WinLose}.
In short, they generally fail to agree with observations.
In terms of metallicity, Cloudy-LZ-LE is the best because it has central metallicities
that are in relatively good agreement with observations.
That said, this calculation, like all the other simulations, shows too steep a gradient in the metal profile at larger radii.
With regards to star formation rate, none of our simulations accurately capture the star formation
history one expects from galaxy clusters:
the overall rate of star formation is too low,
and peaks at too late of a time,
compared to the rates inferred from observations of cluster galaxies.

Notwithstanding the qualitative assessments above and in the table, some of our results are encouraging.
The inclusion of metallicity-dependent cooling rates significantly improves upon previous
work that uses a cooling table that assumes a fixed, relatively high metallicity, and we show
that it is possible to adjust simulations to achieve better specific agreement with observations.
Clearly, a metal-dependent cooling method, like the method of \citet{2008MNRAS.385.1443S, 2011ApJ...731....6S},
is essential to prevent early overcooling of unenriched gas.
Using a distributed feedback model, rather than depositing all of the thermal energy
from supernovae into a single cell, does not have as large of an effect on cluster properties as cooling.
In addition, we find that while metallicity profiles as a function of radius can be altered somewhat based
on choices of feedback parameters, the metallicity profiles never resemble cluster observations.
This points to the need for a more sophisticated feedback model that may include AGN feedback.
Taken as a whole, these results begin to indicate a way forward that uses metallicity-dependent cooling,
and which also addresses the shortcomings of the star formation and feedback model employed in this paper.

We conclude that our current methodologies are not capable of reproducing all observed properties of real clusters.
Therefore, the next step on the road to more realistic clusters is to employ new physical models
that address the shortcomings discussed in this paper.
In particular, it is clear that we must address the unrealistic star formation histories and
metallicity profiles of the clusters.
Without improvements to both, it will be impossible to simulate
the energetics of heating and cooling of observed clusters.

The method we believe will provide substantial improvements to our current methods are ``galaxy particles''
\citep[also called ``galaxy constructs'', or ``galcons,''][]{2008ApJ...683L.111A, 2010ApJ...716..918A}.
Galaxy particles employ a model where the the time evolution of mass, metal, and energy feedback of galaxies are
derived from the observed globally-averaged star formation rate,
and the feedback quantities are deposited into the intergalactic medium isotropically and
over a sphere whose radius is substantially larger than the cell size of the simulation,
thus potentially avoiding the over-cooling.
\citet{2010ApJ...716..918A} compares a cluster simulation employing galaxy
particles to one similar to RS-Single (in particular, in terms of cooling and feedback models).
They find that galaxy particles are more effective at depositing
mass, energy, and metals into the intracluster medium which results in more realistic
temperature, density, metallicity, and entropy profiles.
They find that galaxy particles help reduce the amount of subhalo ``over-merging'' in clusters,
which reduces the central dominance of the simulated cluster.
An additional advantage of such a technique is that it will allow us to include,
in a parametric way, more complex physics such as AGN feedback and merger-driven starburst events.
If successful, one could then use a wider range of cluster observables
(e.g., the Sunyaev-Zel'dovich effect or L$_X$-T relationship) to further constrain calculations.
Galaxy particles are presently being worked on by the \Enzo\ development community,
and we will present the results of cluster simulations using galaxy particles in a future paper.

\acknowledgments

This work was funded by National Science Foundation
(NSF) grant AST 1106437 to J.O.B.
S.W.S. has been supported by a DOE Computational
Science Graduate Fellowship under grant number DE-FG02-97ER25308.
This work utilized the Janus supercomputer, which is supported
by the NSF (award number CNS-0821794)
and the University of Colorado, Boulder.
The Janus supercomputer is a joint effort of the University of Colorado Boulder,
the University of Colorado Denver,
and the National Center for Atmospheric Research.
B.W.O. has been supported in part by grants from the NASA ATFP program
(NNX09AD80G and NNX12AC98G) and through MSU's Institute for Cyber-Enabled Research.
We would like to thank the referee for their helpful comments.

\bibliographystyle{apj}
\bibliography{refs}
\end{document}